%% file: smoothfdr_arxiv.tex
\documentclass[11pt,]{article} 

\usepackage[left=1.25in,top=1.0in,right=1.25in,bottom=1.0in]{geometry}
\usepackage{amsmath,amssymb,amsthm,amsfonts}
\usepackage{caption,subcaption}
\usepackage{times}
\usepackage{xr}
\usepackage{hyperref}
\usepackage{url}
\usepackage{amssymb}
\usepackage[round]{natbib}
\usepackage{amsbsy}
\usepackage{amsthm}
\usepackage{amsmath}
\usepackage{paralist}
\usepackage{bm}
\usepackage{booktabs}
\usepackage{rotating}
\usepackage{cases}
\usepackage{mathtools}
\usepackage{graphicx}
\usepackage{algorithm}
\usepackage{algpseudocode}
\usepackage{setspace}
\usepackage{textcomp}

\newtheorem{theorem}{Theorem}

\newcommand{\N}{\mbox{N}}

\newcommand{\FDRL}{$\text{FDR}_L$ }
\newcommand{\R}{\mathbb{R}}
\newcommand{\vnorm}[1]{\left|\left|#1\right|\right|}

\title{False discovery rate smoothing}

\author{
Wesley Tansey\footnote{Department of Computer Science, University of Texas at Austin, \texttt{tansey@cs.utexas.edu}} \\
Oluwasanmi Koyejo\footnote{Department of Computer Science, University of Illinois at Urbana-Champaign, \texttt{sanmi@illinois.edu}} \\
Russell A. Poldrack\footnote{Department of Psychology, Stanford University, \texttt{poldrack@stanford.edu}} \\
James G.~Scott\footnote{Department of Information, Risk, and Operations Management; Department of Statistics and Data Sciences; University of Texas at Austin, \texttt{james.scott@mccombs.utexas.edu} (corresponding author)}
}

\begin{document}

\maketitle

\begin{abstract}
We present false discovery rate smoothing, an empirical-Bayes method for exploiting spatial structure in large multiple-testing problems.  FDR smoothing automatically finds spatially localized regions of significant test statistics.  It then relaxes the threshold of statistical significance within these regions, and tightens it elsewhere, in a manner that controls the overall false-discovery rate at a given level.  This results in increased power and cleaner spatial separation of signals from noise.  The approach requires solving a non-standard high-dimensional optimization problem, for which an efficient augmented-Lagrangian algorithm is presented.  In simulation studies, FDR smoothing exhibits state-of-the-art performance at modest computational cost.  In particular, it is shown to be far more robust than existing methods for spatially dependent multiple testing.  We also apply the method to a data set from an fMRI experiment on spatial working memory, where it detects patterns that are much more biologically plausible than those detected by standard FDR-controlling methods.  All code for FDR smoothing is publicly available in Python and R.\footnote{\url{https://github.com/tansey/smoothfdr}}

\end{abstract}

\begin{spacing}{1.7}

\input{introduction}

\input{basic_approach}

\input{examples}

\section{Details of model fitting}
\label{sec:model_fitting}
\input{model_fitting}
\input{null_and_alternative}

\input{em_algorithm}
\input{gfl_algorithm}

\input{solution_path}

\section{Simulation experiments}
\label{sec:experiments}
\input{experiments}

\section{Discussion}

\input{discussion}

\end{spacing}

\begin{small}
\bibliographystyle{abbrvnat}
\bibliography{fdrsmooth_bib}
\end{small}

\begin{spacing}{1.7}

\appendix

\input{fmri_data}

\input{plateaus}

\input{benchmark_setup}

\input{fdrregression}

\input{hmrf}

\input{correlated_noise}

\end{spacing}

\end{document}

%% file: introduction.tex
\section{Introduction}

\subsection{Spatial smoothing in the two-groups model}

The traditional problem of multiple testing concerns a group of related null hypotheses $h_1, \ldots, h_n$ that are to be tested simultaneously.  In the simplest form of the problem, a summary statistic $z_i$ is observed for each test.  The goal is to decide which $z_i$ are signals ($h_i = 1$) and which are null ($h_i = 0$) while ensuring that some maximal error rate is not exceeded.  Standard approaches to multiple testing include Bonferroni correction, which controls the overall probability of one or more false positives; and the Benjamini-Hochberg procedure, which controls the false discovery rate, or the expected fraction of false positives among all discoveries \citep{benjamini1995}.  These techniques have been successfully applied across many fields of science, most notably in the analysis of DNA microarrays and other sources of genomic data.

But many large-scale multiple testing problems exhibit spatial patterns that traditional methods do not account for.  Examples include:
\begin{inparaenum}[(1)]
\item fMRI studies, where significant test statistics cluster in anatomically relevant regions of the brain;
\item studies of allele frequencies in biological populations, where genetic loci correspond to physical locations on the chromosome;
\item studies of variation in DNA methylation fraction across specific genomic regions;
\item neural spike-train data recorded from a multi-electrode array, in which electrodes fall at known locations on a two-dimensional lattice; and
\item environmental sensor networks designed to detect spatially localized anomalies.
\end{inparaenum} 

This paper presents a new method called \textit{false discovery rate smoothing} that can learn and exploit the underlying spatial structures in these multiple-testing problems.  FDR smoothing finds spatially localized regions of significant test statistics by solving a specific optimization problem involving the $\ell_1$ penalty.  It then relaxes the threshold of statistical significance within these regions in a manner that controls the global false-discovery rate at a specified level.  

Throughout the paper, we emphasize three key advantages of FDR smoothing:
\begin{description}
\item[Local adaptivity.] FDR smoothing finds localized spatial structure ``out of the box,'' with automated hyperparameter tuning.  This results in increased power and cleaner spatial separation of signals from noise versus standard methods.
\item[Robustness.]  We highlight the failure modes of existing techniques for spatially dependent multiple testing and show that, in contrast, FDR smoothing is much more robust.
\item[Computational efficiency.] FDR smoothing utilizes modern techniques for convex optimization, yielding modest runtimes even for very large spatial problems with $10^6$ observations or more.
\end{description}

\subsection{Connections with existing work}

Our approach incorporates spatial smoothing into the ``two-groups'' model, a popular empirical-Bayes approach for controlling the false-discovery rate that has been advocated by Bradley Efron and many others \citep[e.g.][]{scottberger06,efron:2008,bogdan:etal:2008,scottberger2007,martin:tokdar:2012}.  This strategy leads to a non-standard high-dimensional optimization problem.  To solve this problem, we exploit recent advances in convex optimization for objective functions involving composite regularizers \citep[e.g.~the fused lasso of][]{tibs:fusedlasso:2005}.  \citet{efron:2008} provides a recent review on multiple testing under the two-groups model, while \citet{tibs:taylor:2011} describe a wide class of composite regularizers which they call ``generalized lasso'' problems.  We recommend these two papers to readers who wish to get a deeper sense of these two areas of the literature.

Many authors have considered the problem of multiple testing when the test statistics have a complicated dependence structure \cite[e.g.][]{leek:storey:2008,clarke:hall:2009}.  In contrast, FDR smoothing explicitly uses known spatial structure to inform the outcome of each test.  It therefore differs significantly from these approaches, whose goal is to yield robust FDR-controlling methods in the presence of arbitrarily strong dependence among the test statistics. 

Specific to fMRI analysis, much effort has gone into improving FDR-based procedures. \citet{perone:2004} assume that test statistics are distributed according to a smooth Gaussian random field and their FDR procedure requires additional approximations for practical computation of the hypothesis sets. \citet{benjamini:2007} propose a method geared toward so-called Region of Interest (ROI) studies, where a partitioning is provided a priori, mapping the test locations into distinct clusters. They then use a robust variogram to estimate correlation of test statistics within the cluster and assume no dependence across cluster boundaries. Furthermore, they assume that the primary focus of the testing procedure is at the cluster level, with individual location testing a secondary concern.

But as very recent work by \citet{eklund_etal:2016} demonstates, many common approaches for cluster-level inference are highly non-robust to deviations from the underlying assumption that brain activity can be described using a Gaussian random field.  Our approach differs: it requires no such assumptions about Gaussianity, and no a priori clustering.  And while it is capable of generating clusters \textit{ex post} via a partitioning effect that arises naturally from the $\ell_1$ penalty, it is fundamentally concerned about maintaining the nominal FDR for voxel-level inference rather than cluster-level inference.  Moreover, our technique generalizes far beyond fMRI studies.

Additionally, \citet{schwartzman:2008} perform a smoothing of the test statistic via local averaging.  But their approach operates under fundamentally different assumptions that ours.  For example, they assume that at each spatial location one observes a vector $x$, and the density is required to have the property $f(\mathbf{x}) = f(-\mathbf{x})$.  In the examples we consider, neither assumption is appropriate.

Our work is most directly comparable to two recent methods for spatially-aware multiple testing, both of which were inspired by neuroimaging analysis.  First, \citet{zhang:etal:2011} proposed the \FDRL procedure, which uses spatially smoothed $p$-values to improve power.  Second, \citet{shu:etal:2015} proposed a hidden Markov random field (HMRF) model, which generalizes previous work on one-dimensional spatially-dependent multiple testing \citep{sun:cai:2009} to the multi-dimensional case.  We benchmark our approach against these methods in Section \ref{sec:experiments}.  

FDR smoothing is also conceptually related to a recent proposal by \citet{scott:kass:etal:2014}, called FDR regression.  The two procedures share the goal of improving statistical power for the multiple-testing problem by leveraging external information.  But they operate in very different domains---regression on covariates for FDR regression, versus spatial smoothing here---and they differ from each other in the same way that any spatial smoothing problem differs from any regression problem.  In principle, it is possible to perform spatial smoothing through regression, by introducing a suitable class of basis functions that ``featurize'' space.  But this is rarely done in applied spatial modeling: it is inelegant, computationally burdensome at scale, sensitive to the choice of basis, and unnecessary. In the Appendix, we benchmark FDR smoothing against a version of FDR regression with spatial features, and show  that the latter is not competitive, despite the additional difficulties it poses.

%% file: basic_approach.tex
\section{FDR smoothing: the basic approach}

\subsection{The two-groups model}
\label{section:twogroups}

The FDR smoothing algorithm builds upon the two-groups model for multiple testing \citep{berry1988,efrontibshirani2001}.  Suppose that we have test statistics $z_1, \ldots, z_n$ arise from the mixture
\begin{equation}
\label{basicmixturemodel}
z \sim c \cdot f_1(z) + (1-c) \cdot f_0(z) \, ,
\end{equation}
where $c \in (0,1)$ is an unknown mixing fraction, and where $f_0$ and $f_1$ describe the null ($h_i=0$) and alternative ($h_i=1$) distributions of the test statistics.  In many scientific applications, the $z_i$'s themselves are often the product of a lengthy pre-processing and modeling pipeline.  For example, in many fMRI applications (including ours), the $z_i$'s arise from a complicated voxel-level regression model, which we do not detail here \cite[see, e.g.][]{poldrack_etal:2011}.

To summarize inferences, we report for each $z_i$ the quantity
\begin{equation}
\label{bayesoracle1}
w_i \equiv \mbox{P}(h_i = 1 \mid z_i) =  \frac{ c \cdot f_1(z_i)}{  c \cdot f_1(z_i) + (1- c) \cdot f_0(z_i) } \, .
\end{equation}
This quantity has both a Bayesian and a frequentist interprtation.  To a Bayesian, $w_i$ is the posterior probability that $z_i$ is a signal \citep[e.g.][]{scottberger06,muller:etal:2006}.  To a frequentist, $1-w_i$ is a local false discovery rate.  The two are related by the following equation, from \citet{efrontibshirani2001}.  Let $Z_1 \subset \{z_1, \ldots, z_N\}$ be any subset of the test statistics. Define the quantity
\begin{equation}
\label{eqn:bayesianFDR}
\mathrm{BFDR}(Z_1) = |Z_1|^{-1} \sum_{i: z_i \in Z_1} (1-w_i ) \, .
\end{equation}
\citet{efrontibshirani2001} called this ``Bayesian FDR,'' and showed that under mild conditions it provides a conservative estimate of frequentist FDR.

An important modeling choice in the two-groups model is how to specify the null hypothesis $f_0$.  \citet{efron:2004} distinguishes two important scenarios: the theoretical null, in which $f_0$ is known; and the empirical null, in which $f_0$ is known only up to its functional form (e.g.~Gaussian), and must be estimated jointly with $f_1$ and $c$.   As we show in Section \ref{sec:null_and_alternative}, FDR smoothing can accommodate either scenario in a simple, modular way.

\subsection{Spatial smoothing}
\label{sec:spatial_smoothing}

\paragraph{Formulating the model.}  FDR smoothing involves a conceptually simple modification of the two-groups model (\ref{basicmixturemodel}) that leads to a non-standard high-dimensional optimization problem.  In this section, we provide a high-level overview of the model.  Technical details are deferred to Section \ref{sec:model_fitting}.

Let each $z_i$ be associated with a vertex $s_i \in \mathcal{S}$ in an undirected graph $\mathcal{G}$ with edge set $\mathcal{E}$.  For example, in an fMRI problem, each $s_i$ is a voxel, and $\mathcal{E}$ is a three-dimensional grid.  Suppose that the prior probability in (\ref{basicmixturemodel}) changes from site to site:
\begin{eqnarray}
z_i &\sim& c_i \cdot f_1(z_i) + (1-c_i) \cdot f_0(z_i) \label{eqn:fdrr1} \\
c_i &=& \frac{e^{\beta_i}}{1+e^{\beta_i}} \label{eqn:fdrr2} \, .
\end{eqnarray}
Thus $e^{\beta_i}$ is the prior odds that site $s_i$ has a signal.  Section \ref{sec:model_fitting} describes how $f_0$ and $f_1$ are estimated, but for now we assume that they are fixed.

Let $\boldsymbol\beta = (\beta_1, \ldots, \beta_n)^T$ be the vector of log odds, and let $l(\boldsymbol\beta)$ be the negative log likelihood with $f_0$ and $f_1$ fixed:
$$
l(\boldsymbol\beta) = - \sum_{i=1}^n \log \left[ \left( \frac{e^{\beta_i}}{1+e^{\beta_i}} \right) f_1(z_i) + \left( 1 - \frac{e^{\beta_i}}{1+e^{\beta_i} }\right) f_0(z_i) \right] \, .
$$
We estimate $\boldsymbol\beta$ by penalizing the likelihood: that is, by minimizing the function $f(\boldsymbol\beta) = l(\boldsymbol\beta) +  \lambda J(\boldsymbol\beta)$ for some $\lambda > 0$ and penalty function $J$.  From a Bayesian perspective, this corresponds to the maximum \textit{a posteriori} (MAP) estimate under the prior distribution $\pi(\boldsymbol\beta) \propto e^{- \lambda J(\boldsymbol\beta)}$.

While there are many reasonable choices for $J(\boldsymbol\beta)$, in this paper we penalize the unweighted total variation of $\boldsymbol\beta$ over the graph $\mathcal{G}$.  This leads to the problem
\begin{equation}
\label{eqn:FLobjective0}
\begin{aligned}
& \underset{\boldsymbol\beta \in \R^n}{\text{minimize}}
& & 
l(\boldsymbol\beta) + \lambda \sum_{(i,j) \in \mathcal{E}} |\beta_i - \beta_j| \, ,
\end{aligned}
\end{equation}
which enforces spatial smoothness by penalizing differences in log odds across edges in the graph, and which is closely related to intrinsic conditional autoregressive (CAR) priors.

This choice is motivated by an analogy with image segmentation in computer vision, a problem domain in which total-variation penalties like that in (\ref{eqn:FLobjective0}) have proven successful and computationally efficient \citep[e.g.][]{rudin:osher:faterni:1992}.  In our setup, $\boldsymbol\beta$ represents an ``image'' of prior log odds, and an ``image segment'' corresponds to a region of interest where the proportion of signals is elevated versus the background.  The key difference is that the object being smoothed ($\boldsymbol\beta$) parametrizes a set of weights in a mixture model, rather than a set of means for each pixel in an image.\footnote{To see why ordinary image segmentation does not address the multiple-testing problem, see the examples in Section \ref{subsec:toy_example}.  Total-variation denoising of the $z$-scores can easily find the region of interest for Example 1, but will fail badly for Example 2 (where the signals are not mean-shifted compared to the null).}

Following \citet{tibs:taylor:2011}, we rewrite (\ref{eqn:FLobjective0}) in the following way.  Let $m = |\mathcal{E}|$ be the size of the edge set, and let $D$ be the oriented adjacency matrix of the graph $\mathcal{G}$, which is the $m \times n$ matrix defined as follows.  If $(j,k), j<k$ is the $i$th edge in $\mathcal{E}$, then the $i$th row of D has a $1$ in position $j$, a $-1$ in position $k$, and a $0$ everywhere else.  Thus the vector $D \boldsymbol\beta$ encodes the set of pairwise first differences between adjacent sites in the log odds of being a signal, and $\sum_{(i,j) \in \mathcal{E}} |\beta_i - \beta_j| = \Vert D \boldsymbol\beta \Vert_1$.  We can therefore express the optimization problem as
\begin{equation}
\label{eqn:FLobjective1}
\begin{aligned}
& \underset{\boldsymbol\beta \in \R^n}{\text{minimize}}
& & 
l(\boldsymbol\beta) + \lambda \Vert D\boldsymbol\beta \Vert_1 \, .
\end{aligned}
\end{equation}
This clarifies that the penalty is a composition of two functions applied to $\boldsymbol\beta$: a linear transformation composed with the $\ell_1$ norm.

\paragraph{Using the solution.}
The solution to this optimization problem yields an estimate $\hat{\boldsymbol\beta}$ for the log odds at all sites.  Because the $\ell_1$ penalty encourages sparsity in the first differences of $\hat{\boldsymbol\beta}$ across the edges of the graph, the estimate will partition the nodes of the graph into regions where the log odds are locally constant.  Thus the solution to (\ref{eqn:FLobjective0}) builds spatial structure directly into the estimated prior probabilities $\hat{c}_i$ in Equations (\ref{eqn:fdrr1}) and (\ref{eqn:fdrr2}).  These site-dependent prior probabilities are then used to compute the posterior probability
\begin{equation}
\label{bayesoracle2}
 \hat{w}_i = \mbox{P}(h_i = 1 \mid z_i) =  \frac{ \hat{c}_i \cdot f_1(z_i)}{  \hat{c}_i \cdot f_1(z_i) + (1- \hat{c}_i) \cdot f_0(z_i) } \, .
\end{equation}
Just as with the output of the ordinary two-groups model (\ref{bayesoracle1}), we use these posterior probabilities $w_i$ directly in Equation (\ref{eqn:bayesianFDR}) to find the largest set of discoveries $Z_1$ for which $\mathrm{BFDR}(Z_1)$ satisfies the desired false discovery rate.

%% file: examples.tex
\section{Examples}
\label{sec:examples}

In this section we provide a snapshot of FDR smoothing's performance on both a toy problem and on a real fMRI data set from an experiment on spatial working memory. In both examples, we compare FDR smoothing against a different multiple testing strategy in order to highlight its key advantages relative to other methods.  These examples are meant to be illustrative rather than exhaustive; in Section \ref{sec:experiments} we report the results of a more extensive set of simulation experiments that benchmark FDR smoothing against various other methods. 

\subsection{A toy one-dimensional problem}

\label{subsec:toy_example}

We simulated $z$ scores along a 1D grid of sites $s_i \in \{1, \ldots, 5000\}$ from the following model:
\begin{eqnarray*}
z_i &\sim& c_i \cdot N(\mu_1, \sigma^2_1) + (1-c_i) \cdot \N(0, 1) \\
c_i &=& 
\left\{
\begin{array}{ll}
w_{\mathrm{on}} & \mbox{if $s_i \in [2251, 2750]$} \\
w_{\mathrm{off}} & \mbox{otherwise,}
\end{array}
\right.
\end{eqnarray*}
so that $w_{\mathrm{on}}$ is the fraction of sites that are signals in the 500-site region of interest and $w_{\mathrm{off}}$ the fraction of sites that are signals elsewhere.  In most practical problems $w_{\mathrm{off}}$ is rarely 0, if only for the reason that outliers due to technical or experimental artifacts cannot be eliminated entirely.

\begin{figure}
\begin{center}
\begin{tabular}{cc}
\includegraphics[width=3.0in]{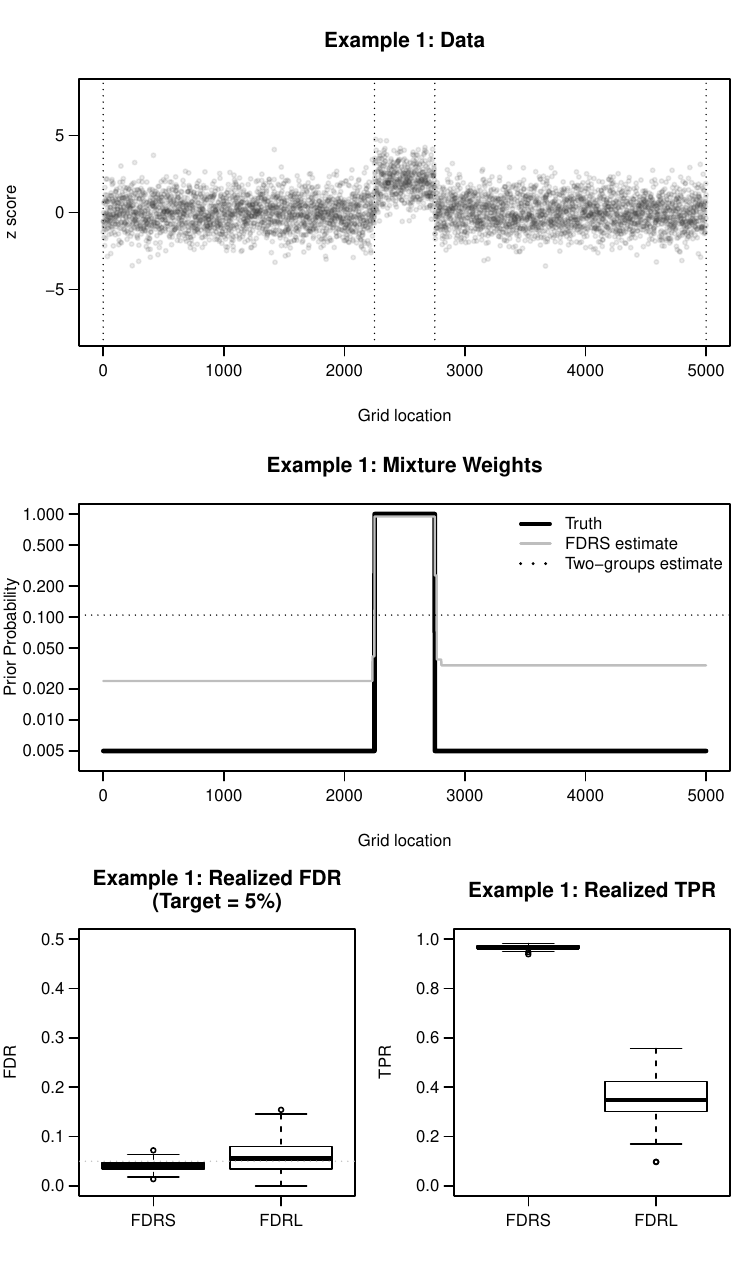} & 
\includegraphics[width=3.0in]{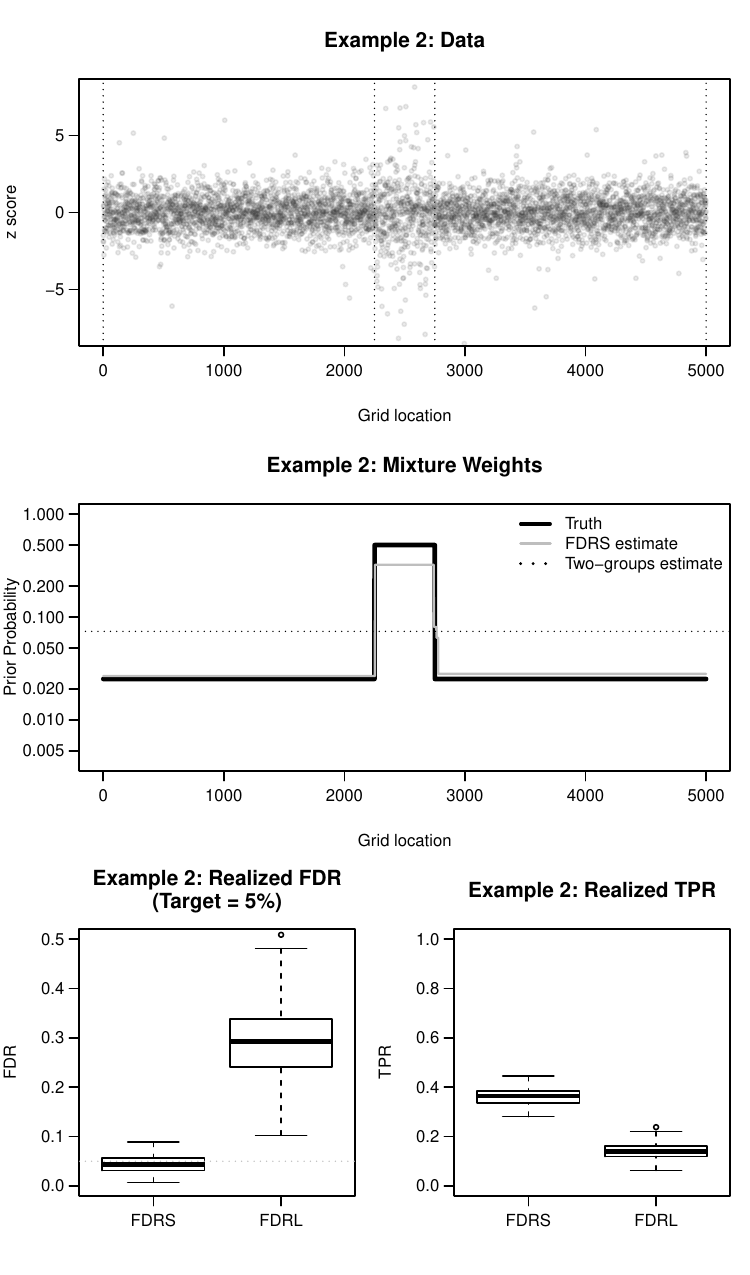}
\end{tabular}
\end{center}
\caption{\label{fig:toy_example_1D} The top two panels show raw $z$ scores from the toy examples of Section \ref{subsec:toy_example} (example 1 left, example 2 right).  The region of interest is sites 2251--2750, shown by vertical dotted lines.  The middle two panels show the true site-level prior probability of a signal, versus the estimates from FDR smoothing (solid grey) and the ordinary two-groups model (dotted).  The bottom four panels show the realized false-discovery rates (FDR) and true-positive rates (TPR) of both FDR smoothing and the \FDRL procedure of \citet{zhang:etal:2011} across 150 simulated data sets.  On example 1, both methods respect the nominal FDR of 5\% on average, although FDR smoothing consistency has higher power.  On example 2, \FDRL violates the nominal FDR yet still has lower power than FDR smoothing (which does obey the nominal FDR).}
\end{figure}

This is a stylized version of a multiple-testing problem that might come up in analyzing allele frequencies or DNA methylation fraction across adjacent sites in the genome \citep[e.g.][]{jaffe:etal:2012}.  Two specific versions of the model were considered:
\begin{description}
\item[Example 1:] $(w_{\mathrm{on}}, w_{\mathrm{off}}) = (1.0, 0.005)$ and $(\mu_1, \sigma^2_1) = (2, 1)$.  This is a relatively easy problem: the $N(2,1)$ signals are well separated from the $N(0,1)$ null hypothesis, the region of interest is pure signal (i.e.~100\% of the $z$ scores there are from $f_1$), and there are almost no signals/outliers outside the region of interest ($w_{\mathrm{off}} = 0.005$).  The top left panel of Figure \ref{fig:toy_example_1D} shows an example data set simulated from this model.
\item[Example 2:] $(w_{\mathrm{on}}, w_{\mathrm{off}}) = (0.5, 0.025)$ and $(\mu_1, \sigma^2_1) = (0, 3^2)$.  This is a much harder problem than the first example.  The signals are overdispersed compared to the null, but also centered at 0; the region of interest is impure signal, since only 50\% of the $z$ scores there are from $f_1$; and there are scattered signals outside the region of interest ($w_{\mathrm{off}} = 0.025$).  In a real example, these scattered signals might be actual genes of interest or simply technical artifacts.  The top right panel of Figure \ref{fig:toy_example_1D} shows an example data set simulated from this model.
\end{description}

Figure \ref{fig:toy_example_1D} shows the results of applying both \FDRL and FDR smoothing (FDRS) to these two examples.  The middle row of two panels shows the true versus reconstructed prior probabilities from FDR smoothing (Example 1 on the left, Example 2 on the right).  The true prior probability $c_i$ as a function of site is shown as a solid black curve, and the FDR smoothing estimate as a grey curve.  For reference, the estimate of the global mixing weight $c$ from the ordinary two-groups model is shown as a dashed line.  Compared with the global estimate, the FDR smoothing estimate shows a favorable blend of adaptability and stability.  For the region of interest (sites 2251--2750), the FDR smoothing estimate of $c_i$ is higher than the global estimate, though not as high as the truth---it is shrunk downwards to the mean.  For all other sites, the estimate is lower than the global estimate, though not as low as the truth---it is shrunk upwards to the mean.

In our simulation studies described in Section \ref{sec:experiments}, FDR smoothing consistently exhibits both of these features: it adapts automatically to spatial patterns in the data, but it also shrinks toward the global mixing weight estimated by the two-groups model.  The site-level adaptation yields improved power.  The shrinkage yields stability, preventing the model from being too aggressive in isolating spurious groups of signals.  

This stability turns out to be a very desirable property in light of the evidence in the bottom four panels of Figure \ref{fig:toy_example_1D}, which show the realized false discovery rate and true positive rate (TPR) across 150 simulated data sets from each of the two examples (bottom left two panels for Example 1, bottom right two panels for Example 2).  For comparison, we also show the realized error rates of the \FDRL procedure from \citet{zhang:etal:2011}.  Several lessons can be drawn from these panels, especially about an important failure mode of \FDRL:
\begin{itemize}
\item FDR smoothing stays below the nominal FDR level of 5\% on both examples.
\item \FDRL stays below the nominal FDR of 5\% on Example 1, albeit with higher variance and lower power (TPR) than FDR smoothing.
\item \FDRL exceeds the nominal FDR on Example 2, yet still has less power than FDR smoothing.
\item Despite these advantages, computation time for FDR smoothing is minimal: roughly 1 second to fit the entire solution path across a grid of $\lambda$ values (see Section \ref{sec:solution_path}).
\end{itemize}
As Section \ref{sec:experiments} will demonstrate more systematically, \FDRL is not robust to situations where the region of interest is not 100\% signal, and where there are outliers outside the region of interest.\footnote{We will see that the HMRF model of \cite{shu:etal:2015} also has this problem.} FDR smoothing, on the other hand, is very robust: the model performs well even under misspecification (see Section \ref{subsec:experiments:misspecification}) and we have only been able to identify one pathological (and easily corrected) scenario where the procedure systematically exceeds the nominal FDR level (see Section \ref{subsec:experiments:pathological}).

\subsection{Finding significant regions in fMRI}

\label{subsec:fmri_example}

\begin{figure}
\begin{center}
\vspace{-0.75in}
\includegraphics[width=0.35\textwidth,trim={6cm 3.8cm 6cm 1cm},clip]{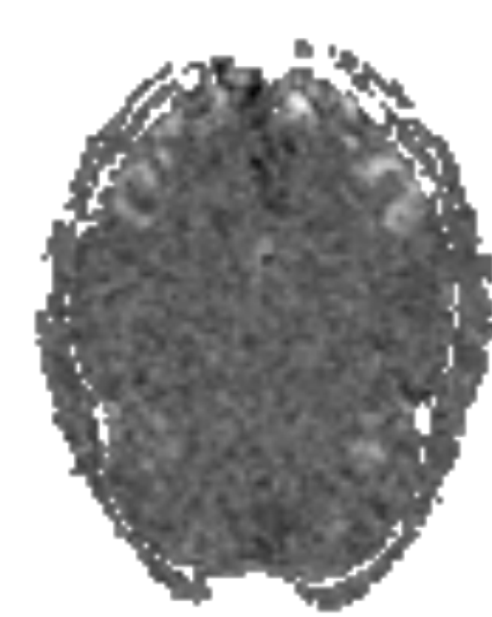}
\includegraphics[width=0.35\textwidth,trim={6cm 3.8cm 6cm 1cm},clip]{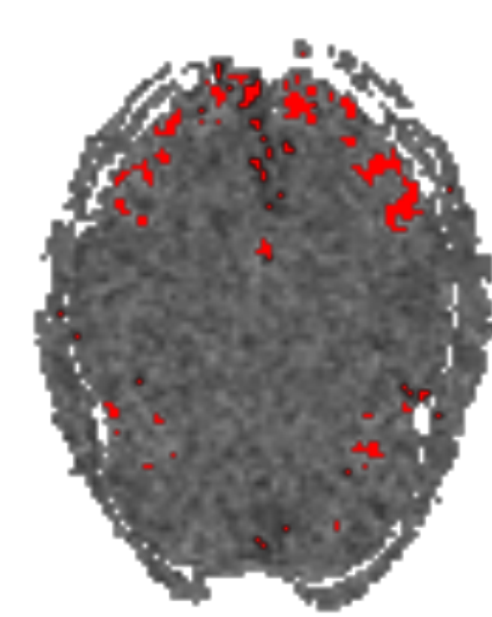}
\includegraphics[width=0.35\textwidth,trim={6cm 3.8cm 6cm 1cm},clip]{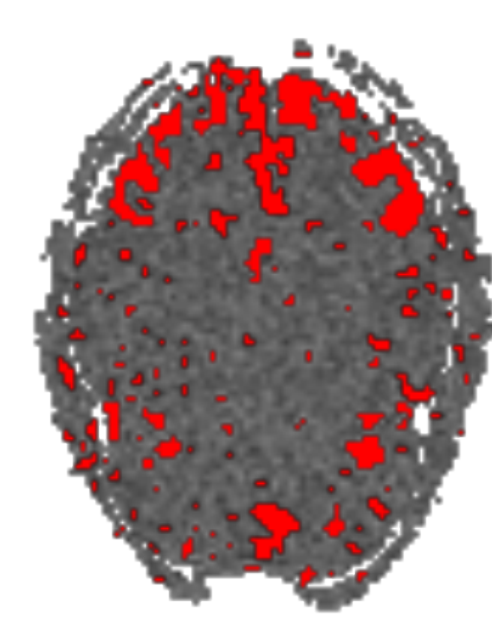}
\includegraphics[width=0.35\textwidth,trim={6cm 3.8cm 6cm 1cm},clip]{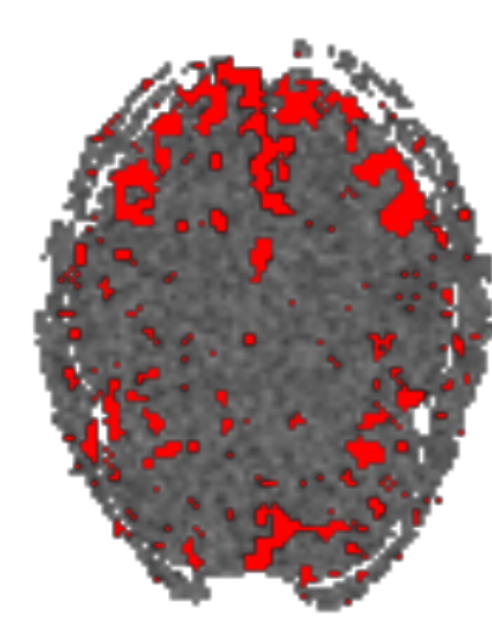}
\includegraphics[width=0.35\textwidth,trim={6cm 3.8cm 6cm 1cm},clip]{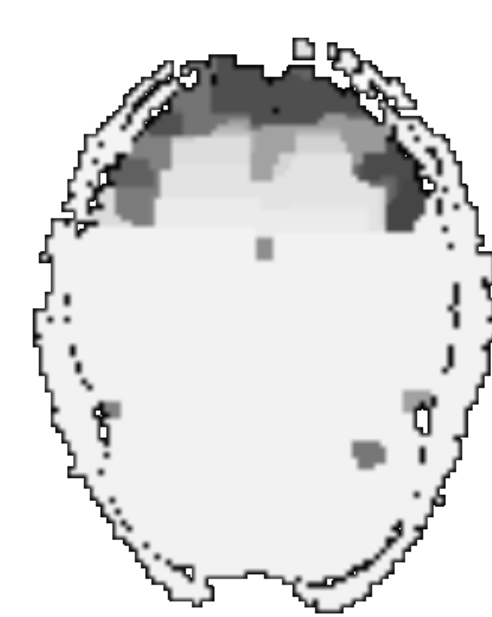}
\includegraphics[width=0.35\textwidth,trim={6cm 3.8cm 6cm 1cm},clip]{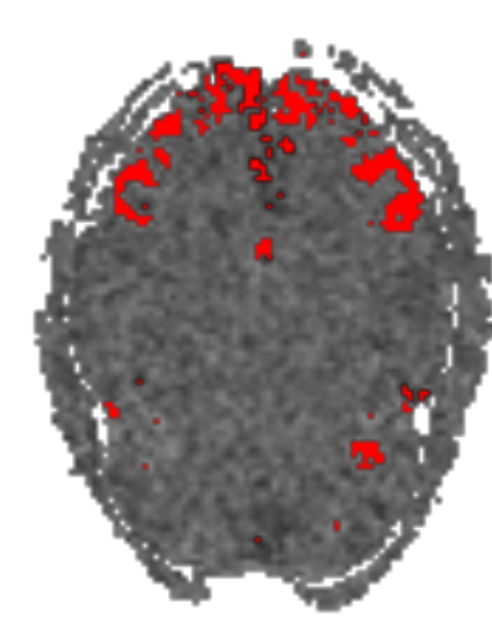}
\end{center}
\caption{\label{fig:russ_example_slice} Top left panel: raw $z$ scores from a horizontal section in an fMRI experiment on spatial working memory.  Darker greys indicate $z$ scores that are larger in absolute value.  There is obvious spatial clustering of the large $z$ scores.  Top right: the significant discoveries that arise from controlling the global false discovery rate at 5\% using the Benjamini-Hochberg procedure. Middle left: the discoveries reported by applying BH to locally-averaged z-values. Middle right: the discoveries reported by the \FDRL algorithm. Bottom left: the spatial pattern estimated by FDR smoothing.  Darker greys correspond to areas of elevated signal density (i.e.~a locally higher fraction of significant z scores). Bottom right panel: discoveries from FDR smoothing at the 5\% level.}
\end{figure}

To show the performance of FDR smoothing in a more realistic scenario, we analyzed data from an fMRI experiment on spatial working memory.  The experiment and analysis protocol are described in detail in Appendix \ref{app:fmri_details}.

The upper left panel of Figure \ref{fig:russ_example_slice} shows an image of $z$ scores, often called the statistical parametric map, arising from the experiment and model.  Signals correspond to regions of the brain that exhibit systematically different levels of activity across the two experimental conditions (difficult versus easy spatial working memory tasks).  The full 3D image has $128 \times 128 \times 75 \approx 1.23$ million voxels.  The left panel of Figure \ref{fig:russ_example_slice} shows a single $128 \times 128$ horizontal section in which the large $z$ scores exhibit an obvious pattern of spatial clustering.  The shapes of these clusters suggest the underlying brain regions associated with the specific cognitive task under study.

The upper-right panel shows the results of applying the Benjamini-Hochberg procedure to these $z$ scores at a 5\% false discovery rate.  The procedure clearly finds regions of adjacent points that are all significant.  However, the edges of these regions are indistinct, and there are many spatially isolated discoveries that presumably represent technical or experiment artifacts.  (It is not scientifically plausible that they are real discoveries for a spatial working memory task.)

The middle two panels show two previous approaches in the literature for leveraging spatial structure. The middle left shows the result of applying local averaging to the $z$ scores and rescaling to unit variance, as suggested for fMRI data by \citet{efron2012}; the middle right shows the output of the \FDRL procedure. Both methods are overly-aggressive in their reporting, likely yielding several more false discoveries than the specified 5\% threshold in the areas away from the top of the image.  We conclude this on the basis of two facts: (1) these two methods exhibit exactly these deficiencies on the simulated examples, in Sections 3.1 and 5;  and (2) more importantly, the extra regions ostensibly ``discovered'' by \FDRL and smoothed BH are not scientifically plausible based on what is known about spatial reasoning.  Our interpretation is that \FDRL (and smoothed BH, which is very similar) are suffering from the some of the non-robustness problems exhibited in Section 3.1

Now consider the bottom two panels.  The bottom left panel shows our procedure's estimated partition of the raw data shown in the left panel.  Darker greys correspond to signal-dense areas containing locally higher fraction of significant $z$ scores; lighter areas correspond to signal-sparse areas containing a locally lower fraction of significant $z$ scores.  The bottom right panel then shows the final output: the discovered signals at the 5\% FDR level.  Compared with the other procedures, the image reveals regions of significant signals that are more biologically plausible, detecting regions in the bilateral prefrontal cortex that are commonly associated with working memory function \citep{owen:etal:2005}. This reflects the local adaptivity of FDR smoothing: it loosens the threshold for significance in the apparently interesting regions and tightens it in the uninteresting regions.

The partitions shown in the right panel of Figure \ref{fig:toy_example_1D} and the bottom left panel of Figure \ref{fig:russ_example_slice} are estimated by our specialized adaptation of an edge-detection algorithm used to denoise images of natural scenes.  However, we emphasize that FDR smoothing is \textit{not} simply denoising the $z$ scores. That is, the estimated partition does not merely pick out areas in which the actual $z$ scores (raw pixel values) are locally constant or locally homoscedastic.  Rather, it picks out areas in which the unknown true fraction of signals is locally constant.  Unlike the pixel values, these local enrichment fractions are not actually observed by the experimenter.  This is a significant complication not present in image denoising and is the fundamental statistical problem addressed by our approach.

%% file: model_fitting.tex

In this section, we describe our model-fitting process in much more detail.
\begin{itemize}
\item Our formulation of the FDR-smoothing problem assumes that both $f_0(z)$ and $f_1(z)$ are known, which is obviously untrue in practice.  Section \ref{sec:null_and_alternative} describes a simple approach for estimating these two distributions that combines techniques from \citet{efron:2004} and \citet{martin:tokdar:2012}.
\item Sections \ref{sec:em_algorithm} and \ref{sec:augmented_lagrangian} describe how the FDR-smoothing optimization problem can be solved efficiently, even for very large graphs.
\item Section \ref{sec:solution_path} describes a path-based procedure for choosing the regularization parameter $\lambda$.
\end{itemize}

%% file: null_and_alternative.tex
\subsection{Estimating the null and alternative densities}
\label{sec:null_and_alternative}

Our overall fitting approach emphasizes modularity.  We estimate the null and alternative densities $f_0$ and $f_1$ separately from $\boldsymbol\beta$, before solving the FDR-smoothing optimization problem (\ref{eqn:FLobjective0}).  We do so by fitting the ordinary two-groups model from Section \ref{section:twogroups}, thereby ignoring the spatial information contained in the graph $\mathcal{G}$.  This gives us estimates for $f_0$ and $f_1$, which we then fix and use to estimate $\boldsymbol\beta$ using the model described in Section \ref{sec:spatial_smoothing}.

This two-stage procedure may sound ad-hoc, but in fact has a simple justification. Consider the underlying site-level mixture model $z_i \sim m_i(z) \equiv c_i f_1(z) + (1-c_i) f_0(z)$, where $c_i$ is the prior probability of a signal at site $i$.  Under this model, the marginal distribution of the $z_i$'s \emph{across sites} is a ``mixture of mixtures.''  Thus if $z$ is a randomly selected test statistic from the data set, then marginally
$$
z \sim  \frac{1}{N} \sum_{i=1}^N m_i(z) \, .
$$
This appears at first glance to be a mixture model with $N$ mixture components, one per site.  But in fact, because each component is itself a mixture of $f_0$ and $f_1$, the marginal for $z$ is actually a mixture with only two components.  These are exactly the components we want to estimate:
\begin{equation}
\label{eqn:mixtureofmixtures}
z \sim  \frac{1}{N} \sum_{i=1}^N m_i(z) = \bar{c} f_1(z) + (1-\bar{c}) f_0(z), 
\end{equation}
where the mixing weight $\bar{c} = \sum c_i /N$ is the average prior probability of a signal across all sites.

This has a striking implication: we can estimate $f_0$ and $f_1$ for our spatially varying model even if we assume that the \emph{ordinary} two-groups model is true---that is, by ignoring spatial information and using the marginal distribution of test statistics alone.  Thus we follow a two-step procedure: estimate $f_0$ and $f_1$ from (\ref{eqn:mixtureofmixtures}) using standard tools, and then fix these to estimate the spatially varying $c_i$.  (The first step also recovers $\bar{c}$, although this information is not used directly.)

We now describe our use of standard tools to estimate $f_0$ and $f_1$ directly from (\ref{eqn:mixtureofmixtures}).  An open question is whether using spatial information would sharpen these estimates; this is plausible, but we leave it for future work.

\paragraph{Estimating $f_0$.}  In some cases, $f_0$ is taken directly from the distributional theory of the test statistic in question (e.g.~standard Gaussian) and therefore need not be estimated at all.  For such problems where a theoretical null describes the data well, this step can be skipped.

However, as \citet{efron:2004} argues, in many multiple-testing problems the data are poorly described by the theoretical null.  In such cases, an empirical null hypothesis with known parametric form but unknown parameters must be estimated in order to produce reasonable results.  For the problems considered in this paper, the null hypothesis is assumed to be a Gaussian with unknown mean and variance: $f_0(z) = \N(z \mid \mu_0, \sigma^2_0)$.  But in principle the same matching technique can be used in any exponential family.

\begin{figure}
\begin{center}
\includegraphics[width=5in]{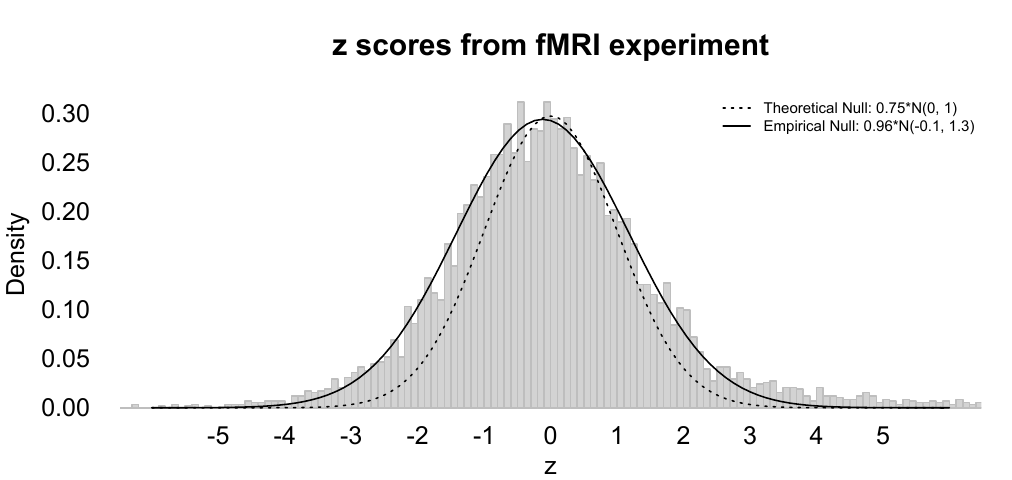}
\caption{\label{fig:empirical_null} Theoretical versus empirical null for the fMRI data from Section \ref{subsec:fmri_example}.}
\end{center}
\end{figure}

To estimate $\mu$ and $\sigma$, we apply the central-matching method of \citet{efron:2004}, which uses the shape of the histogram of test statistics near zero (which come mostly or exclusively from the null distribution).  Specifically, let $Z_C$ be the subset comprising some central fraction (we use 1/3) of the $z$ scores.  Central matching proceeds in three steps:
\begin{enumerate}
\item Construct a smooth estimate $\hat{g}(z)$ of the log density of the test statistics in $Z_C$, and let $z_0$ be the point where $\hat{l}(z)$ obtains its maximum.
\item Form a second-order Taylor approximation of $\hat{g}(z)$ about its maximum:
$$
\hat{g}(z) \approx q(z) = \frac{d_2}{2} (z - z_0)^2 + d_1 (z- z_0) + d_0 \, .
$$
\item The quadratic approximation on the log scale corresponds to a Gaussian on the original scale.  Therefore set the mean and standard deviation of the empirical null using the slope and curvature of $q(z)$:
\begin{eqnarray*}
\mu_0 &=& z_0 - \frac{d_1}{2d_2} \\
\sigma_0 &=& \sqrt{-\frac{1}{2d_2}} \, .
\end{eqnarray*}
\end{enumerate}
For an alternative approach to estimating an empirical null, see \citet{martin:tokdar:2012}.

Figure \ref{fig:empirical_null} shows both the theoretical and empirical null for the fMRI data analyzed in Section \ref{subsec:fmri_example}.  This figure indicates that the theoretical null is likely inadequate, and our analysis therefore uses false discovery rates estimated using the empirical null.

\paragraph{Estimating $f_1$.} 

Having fixed an estimate for $f_0(z)$, $f_1(z)$ can be estimated by any of several existing methods for one-dimensional Gaussian deconvolution, including finite mixture models or Dirichlet-process models \citep{domuller2005}.  We use and recommend the predictive-recursion algorithm of \citet{newton:2002} because it is fast, flexible, and enjoys strong guarantees of accuracy \citep[see][]{tokdar:martin:ghosh:2009}.  Predictive recursion generates a nonparametric estimate $\hat f_1(z)$ for the marginal density under the alternative after a small number of passes through the data.\footnote{In our examples we use 50 passes, although in our experience 10 passes is virtually always sufficient to yield stable estimates.}  For further details, see \citet{martin:tokdar:2012}; for pseudo-code, see \citet{scott:kass:etal:2014}.

%% file: em_algorithm.tex
\subsection{An expectation-maximization algorithm}
\label{sec:em_algorithm}

We now turn to the details of solving the optimization problem in (\ref{eqn:FLobjective0}).  This problem is hard for at least two reasons: the likelihood term $l(\boldsymbol\beta)$ is nonconvex, and the penalty term is nonseparable in $\boldsymbol\beta$.  We are not aware of any algorithm that is guaranteed to find the global minimum efficiently, even for fixed $\lambda$, and the method we describe below finds only a local minimum.  Nonetheless, this paper marshals evidence that the local solutions actually found by our algorthm yield good reconstructions of underlying spatial patterns and better power than existing FDR-controlling methods.

We handle the likelihood term with a simple data-augmentation step that leads to an expectation-maximization (EM) algorithm.  For now, we assume that $\lambda$ is fixed; we describe our method for choosing this hyperparameter in Section \ref{sec:solution_path}.  Introduce binary latent variables $h_i$ such that
\begin{eqnarray*}
z_i &\sim& 
\left\{ \begin{array}{l l}
f_1(z_i) & \mbox{if \ } h_i = 1 \, , \\
f_0(z_i) & \mbox{if \ } h_i = 0 \\
\end{array}
\right. \\
\mbox{P}(h_i = 1) &=&  \frac{e^{\beta_i}}{1+e^{\beta_i}} \, .
\end{eqnarray*}
Marginalizing out the $h_i$ clearly gives us the original model (\ref{eqn:fdrr1}).  Treating $\boldsymbol h$ as fixed gives the complete-data negative log likelihood:
\begin{equation}
\label{eqn:completedataloglike}
l(\boldsymbol\beta, \boldsymbol h) = \sum_{i=1}^n \left\{  \log \left(1 + e^{\beta_i}\right) - h_i \beta_i \right\} \, .
\end{equation}
With $\boldsymbol h$ fixed, this is a convex function in $\boldsymbol\beta$ and is equivalent to the negative log likelihood of a logistic-regression model with identity design matrix.

Therefore, a stationary point of (\ref{eqn:FLobjective1}) may be found via a conceptually simple EM algorithm.  Suppose that the step-$k$ estimate for the underlying image of log odds is $\boldsymbol\beta^{(k)}$.  In the E step, we compute $q^{(k)}(\boldsymbol\beta) = E\{ l(\boldsymbol\beta, \boldsymbol h) \mid \boldsymbol\beta^{(k)} \}$.  Because the complete-data log likelihood is separable and linear in the $h_i$, we simply plug in the conditional expected value for $h_i$, given the current guess for $\beta_i$, into $l(\boldsymbol\beta, \boldsymbol h)$.  Since $h_i$ is a binary random variable, this is just the conditional probability that $h_i = 1$:
\begin{equation}
w_i^{(k)} = E(h_i \mid \boldsymbol\beta^{(k)}, z_i) = \frac{c^{(k)}_i \cdot f_1(z_i)}{ c^{(k)}_i \cdot f_1(z_i) + (1-c^{(k)}_i) \cdot f_0(z_i) } \label{bayesoracle3} \, ,
\end{equation}
where $c^{(k)}_i$ is the prior probability that site $i$ produces a signal, given by the inverse logit transform of the current estimate $\beta_i^{(k)}$ from equation (\ref{eqn:fdrr2}).

In the M step, we maximize the complete-data log likelihood.  This requires solving the convex sub-problem
\begin{equation}
\label{eqn:completedata_problem}
\begin{aligned}
& \underset{\boldsymbol\beta \in \R^n}{\text{minimize}}
& & 
\sum_{i=1}^n \left\{  \log \left(1 + e^{\beta_i}\right) - w_i \beta_i \right\}  + \lambda \Vert D \boldsymbol\beta \Vert_1 \, ,
\end{aligned}
\end{equation}
 where the $w_i$ are the complete-data sufficient statistics.

To solve this sub problem, we expand $l(\boldsymbol\beta, \boldsymbol w)$ in a second-order Taylor approximation at the current iterate $\boldsymbol x$.  This turns the $M$ step into a weighted least-squares problem with a generalized-lasso penalty.  Thus up to a constant term not depending on $\boldsymbol\beta$, the intermediate problem to be solved is,
\begin{equation}
\label{eqn:FLobjective2}
\begin{aligned}
& \underset{\boldsymbol\beta \in \R^n}{\text{minimize}}
& & 
\left[ \nabla \ l(\boldsymbol x, \boldsymbol w) \right]^T (\boldsymbol\beta - \boldsymbol x) + \frac{1}{2} (\boldsymbol\beta - \boldsymbol x)^T H(\boldsymbol x,\boldsymbol w) (\boldsymbol\beta - \boldsymbol x) + \lambda \Vert D \boldsymbol\beta \Vert_1 \, ,
\end{aligned}
\end{equation}
where $\nabla \ l(\boldsymbol x, \boldsymbol w)^T$ and $H(\boldsymbol x, \boldsymbol w)$ are the gradient and Hessian with respect to the first argument of the complete-data negative log likelihood $l(\boldsymbol\beta, \boldsymbol w)$, evaluated at the current iterate $\boldsymbol\beta^{(k)}$ (denoted generically as $\boldsymbol x$).  These are simple to evaluate:
\begin{eqnarray*}
\left[ \nabla \ l(\boldsymbol x, \boldsymbol w) \right]_i &=& \frac{e^{x_i}}{1 + e^{x_i}} - w_i \\
\left[ H(\boldsymbol x, \boldsymbol w) \right]_{i,j} &=& \left\{ \begin{array}{l l}
\frac{e^{x_i}}{(1+e^{x_i})^2} & \mbox{if \ } i = j \, \\
0 & \mbox{if \ } i \neq j 
\end{array}
\right. 
\end{eqnarray*}

The Hessian matrix is diagonal because the log likelihood is separable in $\beta_i$. Ignoring terms that are constant in $\boldsymbol\beta$, the solution to (\ref{eqn:FLobjective2}) can be expressed as the solution of a penalized, weighted least-squares problem:
\begin{eqnarray}
\label{eqn:leastsquaresderiv}
\hat{\boldsymbol\beta} = \arg \min_{\boldsymbol\beta \in \R^n} \left\{  \sum_{i=1}^n \frac{\eta_i(y_i - \beta_i)^2}{2} + \lambda \Vert D \boldsymbol\beta \Vert_1  \right\} \, ,
\end{eqnarray}
with working responses $y_i$ and weights $\eta_i$ given as follows:
\begin{eqnarray*}
y_i &=& x_i - \frac{c_i - w_i}{\eta_i} \\ 
\eta_i &=& c_i (1-c_i) \\
c_i &=& \frac{e^{x_i} }{1 + e^x_i} \, .
\end{eqnarray*}
Recall that $\boldsymbol x$ is the point at which the Taylor expansion for the complete-data log likelihood is computed.  In our EM algorithm, this is the current estimate $\boldsymbol\beta^{(k)}$.

Thus the overall steps of algorithm can be summarized as follows.
\begin{description}
\item[1) E-step:] Use formula (\ref{bayesoracle3}) to form the complete-data sufficient statistics $w_i$, given the current estimate of $\boldsymbol\beta$, to get the complete-data negative log likelihood $l(\boldsymbol\beta, \boldsymbol w)$ in (\ref{eqn:completedataloglike}).
\item[2) Quadratic approximation:] Expand $l(\boldsymbol\beta, \boldsymbol w)$ in a second-order Taylor series about the current iterate $\boldsymbol x \equiv \boldsymbol\beta^{(k)}$, thereby forming the ``quadratic + penalty'' surrogate sub-problem in (\ref{eqn:leastsquaresderiv}).
\item[3) Penalized weighted least squares:] Solve the surrogate problem (\ref{eqn:leastsquaresderiv}) using the augmented-Lagrangian method described in Section \ref{sec:augmented_lagrangian}.
\end{description}
In principle, a full M step requires that steps 2 and 3 be interated until local convergence after each E step. In practice, we take a partial M step by iterating steps 2 and 3 only once.  This speeds the algorithm up: step 3 is by far the most computationally expensive, and we want it to be using sufficient statistics $w_i$ that are as up-to-date as possible.  Moreover, as long as the complete-data objective function is improved at each step, the resulting sequence of iterates still converges to a stationary point of (\ref{eqn:FLobjective1}).

%% file: gfl_algorithm.tex
\subsection{Solving the M-step via graph-based TV denoising}
\label{sec:augmented_lagrangian}

The most computationally expensive part of the FDR smoothing algorithm is the need to repeatedly solve the graph-fused lasso problem in \eqref{eqn:leastsquaresderiv}. Even though the GFL is convex, the massive size and graph structure of fMRI scans make classical techniques inefficient. Solving \eqref{eqn:leastsquaresderiv} therefore requires a highly efficient, scalable method.

Many special cases of this problem have been studied in the literature, each with highly efficient, specialized solutions. For example, when $\mathcal{G}$ is a one-dimensional chain graph, \eqref{eqn:leastsquaresderiv} is the ordinary (1D) fused lasso \citep{tibs:fusedlasso:2005}, solvable in linear time via dynamic programming \citep{johnson:2013}. When $\mathcal{G}$ is a D-dimensional grid graph, \eqref{eqn:leastsquaresderiv} is often referred to as total-variation denoising \citep{rudin:osher:faterni:1992}, for which several efficient solutions have been proposed \citep{chambolle:darbon:2009,barbero:sra:2011,barbero:sra:2014}. In contrast to these methods, we next develop a method to solve the GFL over a general graph.


The core idea of our algorithm is to decompose a graph into a set of trails. We note two preliminaries. First, every graph has an even number of odd-degree vertices \citep{west:2001}. Second, if $\mathcal{G}$ is not connected, then the objective function is separable across the connected components of $\mathcal{G}$, each of which can be solved independently. Therefore, for the rest of the section we assume that the edge set $\mathcal{E}$ forms a connected graph.

We also remind the reader of some basic terminology in graph theory. A \textit{walk} is a sequence of vertices, where there exists an edge between the preceding and following vertices in the sequence. A \textit{trail} is a walk in which all the edges are distinct. An \textit{Eulerian trail} is a trail which visits every edge in a graph exactly once. A \textit{tour} (also called a \textit{circuit}) is a trail that begins and ends at the same vertex. An \textit{Eulerian tour} is a circuit where all edges in the graph are visited exactly once.

The following theorem from \citet{west:2001} states that any connected graph can be decomposed into a set of trails T on which our optimization algorithm can operate.

\begin{theorem}\label{thm:2ktrails} \citep[Thm 1.2.33]{west:2001}. The edges of a connected graph with exactly $2k$ odd-degree vertices can be partitioned into $k$ trails if $k > 0$. If $k = 0$, there is an Eulerian tour. Furthermore, the minimum number of trails that can partition the graph is $max(1, k)$.
\end{theorem}

This theorem enables us to rewrite the penalty term in \eqref{eqn:leastsquaresderiv} as a summation over trails $\mathcal{T} = \{t_1, \ldots, t_k\}$, where each trail contains the list of (start, end) vertices for each edge in its walk:
\begin{equation}
\label{eqn:trails_objective}
\begin{aligned}
& \underset{\boldsymbol\beta \in \R^p}{\text{minimize}}
& & 
\sum_{i=1}^n \frac{\eta_i(y_i - x_i)^2}{2} + \lambda \sum_{t \in \mathcal{T}} \sum_{(r,s) \in t} |\beta_r - \beta_s| \, .
\end{aligned}
\end{equation}
Denoting the weighted squared loss portion of the objective as $\ell(\mathbf{y}, \boldsymbol\beta)$, slack variables $\mathbf{z}$ are then introduced for each $\beta_i$ in the penalty term, which results in the following equivalent problem:
\begin{equation}
\label{eqn:trails_objective2}
\begin{aligned}
& \underset{\boldsymbol\beta \in \R^n}{\text{minimize}}
& & 
\ell(\mathbf{y}, \boldsymbol\beta) + \lambda \sum_{t \in \mathcal{T}} \sum_{(r,s) \in t} |z_r^{(t)} - z_s^{(t)}|\, . \\
& \text{subject to}
& & z_r^{(t)} = \beta_r \\
& & & z_s^{(t)} = \beta_s \, ,
\end{aligned}
\end{equation}
where the constraints hold for all pairs $(r,s) \in t$, for all $t \in \mathcal{T}$.  We can then solve this problem efficiently via an ADMM routine \citep{boyd:etal:2011} with the following updates:
\begin{align}
\label{eqn:admm_updates_beta}
\hat{\boldsymbol\beta}^{(j+1)} & = \underset{\boldsymbol\beta}{\text{argmin}} \left( \ell(\mathbf{y}, \boldsymbol\beta) + \frac{\alpha}{2}\vnorm{A\boldsymbol\beta - \mathbf{z}^{(j)} + \mathbf{u}^{(j)}}^2 \right) \\
\label{eqn:admm_updates_z}
\mathbf{z}^{(t,j+1)} & = \underset{\mathbf{z}}{\text{argmin}} \left( \frac{\alpha}{2} \sum_{r \in t} (\tilde{y}_r - z^{(t)}_r)^2 + \sum_{(r,s) \in t} |z^{(t)}_r - z^{(t)}_s|  \right) \; , \quad t \in \mathcal{T} \\
\label{eqn:admm_updates_u}
\mathbf{u}^{(j+1)}& = \mathbf{u}^{(j)} + A\boldsymbol\beta^{(j+1)} - \mathbf{z}^{(k+1)} \, ,
\end{align}
where $u$ is the scaled dual variable, $\alpha$ is the scalar penalty parameter, $\tilde{y}_r = \beta_r - u_r$, and $A$ is a sparse binary matrix used to encode the appropriate $\beta_i$ for each $z^{(t)}_j$. We point out that solving \eqref{eqn:admm_updates_z} corresponds to solving a weighted 1-dimensional fused lasso problem, which can be done in linear time via an efficient dynamic programming routine \citep{johnson:2013,glmgen:2014}. We iterate the updates in \eqref{eqn:admm_updates_beta}-\eqref{eqn:admm_updates_u} in order until the dual and primal residuals have sufficiently small norms.

This trail-based GFL algorithm is highly efficient and can solve the GFL for an fMRI scan locally on a single machine in seconds. In our preliminary experiments, we found the method to be substantially more efficient than any other general GFL method and nearly as efficient as the specialized methods mentioned earlier for multidimensional grid graphs.  The key update steps in our ADMM algorithm is also embarrassingly parallel: each $\hat{\beta}_i$ in \eqref{eqn:admm_updates_beta} and each $\mathbf{z}^{(t)}$ in \eqref{eqn:admm_updates_z} can be solved independently of the other nodes and trails, respectively. For even more massive problems, where the data may not fit in memory, our algorithm will continue to scale well due to its distributed nature. Finally, we note that our algorithm is extensible to other smooth, convex loss functions $\ell$, though we do not explore such extensions here.  A more detailed presentation of this algorithm is available as a supplemental file, ``A fast and flexible algorithm for the graph-fused lasso.''

%% file: solution_path.tex
\subsection{Choosing the regularization parameter}
\label{sec:solution_path}

Once the null and alternative densities have been estimated, the only remaining tuning parameter in FDR smoothing is $\lambda$, the amount of regularization applied to the vector of first differences in log odds across the edges in $\mathcal{G}$.  We now describe our method for choosing $\lambda$ in a data-adaptive way.

\begin{figure}
\begin{center}
\includegraphics[width=6in]{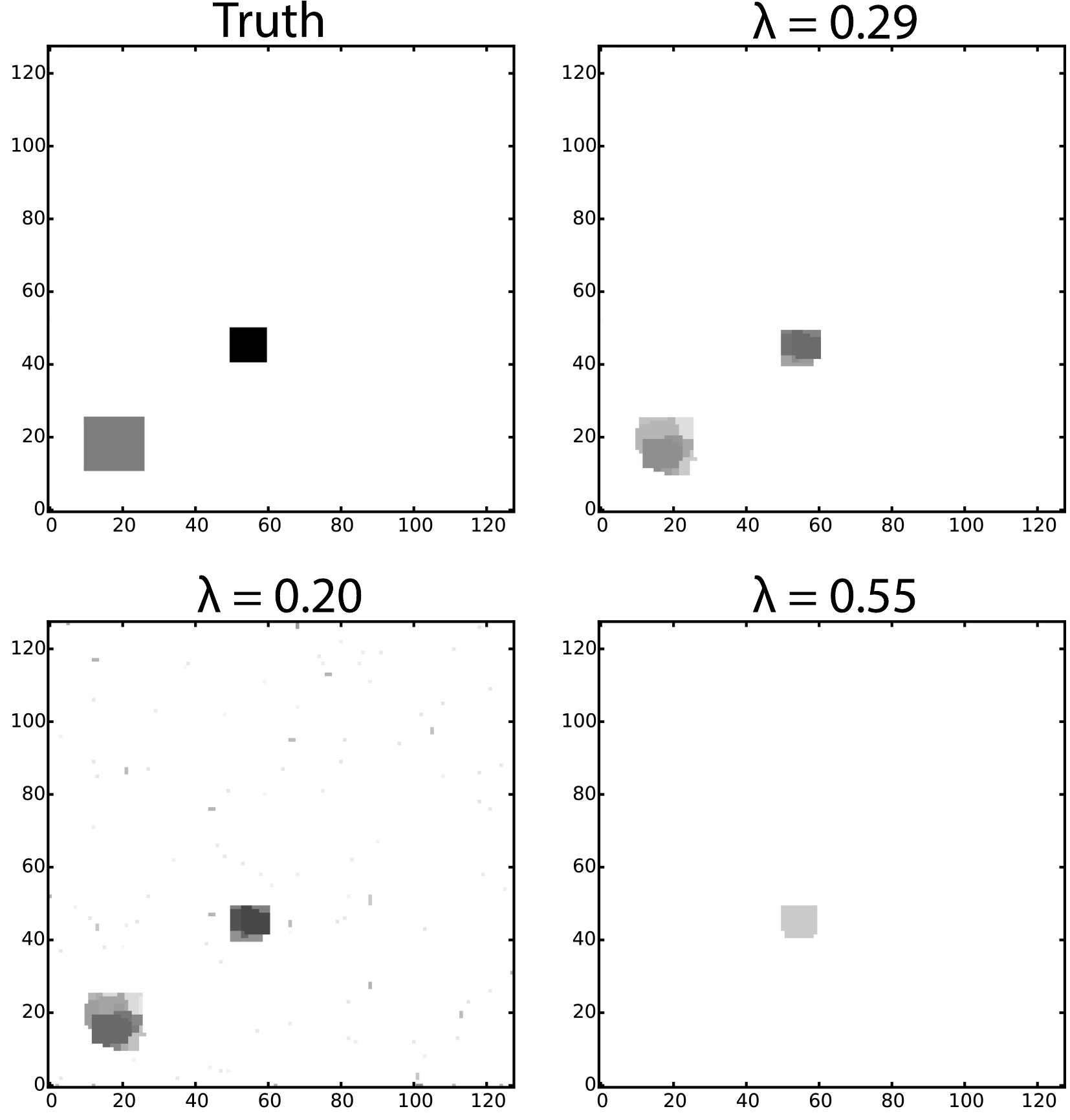}
\end{center}
\caption{\label{fig:lambda_comparisons} Comparisons of different choices of the $\lambda$ penalty parameter. Choosing $\lambda$ too small (bottom left) will produce a grainy reconstruction that overfits the data. Choosing $\lambda$ too large (bottom right) will oversmooth the data and potentially lose crucial structure.  Our path-based method for choosing $\lambda$ results in the choice shown in the top right panel.}
\end{figure}

Figure \ref{fig:lambda_comparisons} illustrates the importance of choosing an appropriate $\lambda$ value.  The top left panel depicts an underlying image of prior probabilities.  We used this image to generate $z$ scores according to the spatially varying two-groups model in equations (\ref{eqn:fdrr1}) and (\ref{eqn:fdrr2}) with a specific choice of $f_1$.  (For details, see the ``small signal'' experiment in Section \ref{sec:experiments}.)  Choosing $\lambda$ too small, as in the bottom left panel, produces a grainy reconstruction that overfits the data. Choosing $\lambda$ too large, as in the bottom right panel, results in oversmoothing and the loss of interesting spatial structure. Our procedure yields the choice of $\lambda$ shown in the top-right panel.  The true regions are recovered with reasonable accuracy, and the graininess of the bottom left panel is avoided.

We avoid having to hand-tune $\lambda$ in an ad-hoc fashion by adopting the following approach, based on the same solution-path idea that is often used to set $\lambda$ in $\ell^1$ problems \citep[e.g.][]{tibs:taylor:2011}.
\begin{enumerate}
\item Calculate the FDR-smoothing solution $\hat{\boldsymbol\beta}(\lambda)$ across a decreasing grid of regularization parameters $\lambda_M > \lambda_{M-1} > \cdots > \lambda_1$, using the solution for $\lambda_s$ as a warm start to find the $\lambda_{s-1}$ solution.
\item For each solution $\hat{\boldsymbol\beta}_s$ corresponding to point $\lambda_s$ on the grid, calculate a relative quality measure $J(\hat{\boldsymbol\beta}_s)$.
\item Choose $\lambda_s$ to be the point in the grid where the quality measure is smallest.
\end{enumerate}
The choice of quality measure should enforce a compromise between the fit and complexity of the reconstructed image.  Perhaps the two most common approaches are AIC and BIC. Let $\phi(\boldsymbol\beta)$ be the degrees of freedom of the estimator and $l(\boldsymbol\beta)$ the maximized value of the log likelihood.  Then up to constants not depending on $\boldsymbol\beta$,
\begin{align}
\label{eqn:aic}
\mathrm{AIC}(\boldsymbol\beta) &= -2\log l(\boldsymbol\beta) + 2\phi(\boldsymbol\beta) \\
\label{eqn:bic}
\mathrm{BIC}(\boldsymbol\beta) &= -2\log l(\boldsymbol\beta) + \log(n) \phi(\boldsymbol\beta) \, .
\end{align} 

For simple one-dimensional problems under squared error loss, calculating the degrees of freedom of the generalized lasso equates to counting the number of change points along the $x$ axis. The two-dimensional extension of this result appeals to Stein's lemma, and involves counting the number of distinct contiguous 2-d regions or \textit{plateaus} in $\boldsymbol\beta$ \citep{tibshirani:taylor:2012}; these results can analogously be extended to arbitary graphs, where a plateau equates to a connected subgraph. For example, the true prior in the upper left panel of Figure \ref{fig:lambda_comparisons} has three plateaus: the two darker squares, and the white background.

Unfortunately, this remarkable result on the degrees of the freedom of the generalized lasso applies only to problems where $l(\boldsymbol\beta)$ is squared error loss.  We are aware of no analogous results in more complicated situations involving mixture models such as ours. Therefore, we cannot plug in the true degrees of freedom when calculating AIC and BIC, because it is not known.  In the absence of a better alternative, we use the number of plateaus as a surrogate for the degrees of freedom.  This is a heuristic solution, but one that seems to yield good performance in practice.  The upshot is that if a good estimator for the true degrees of freedom could be found, it is likely that a smarter $\lambda$ could be chosen automatically, and that our overall method could be improved.

\begin{figure}
\begin{center}
\includegraphics[width=6in]{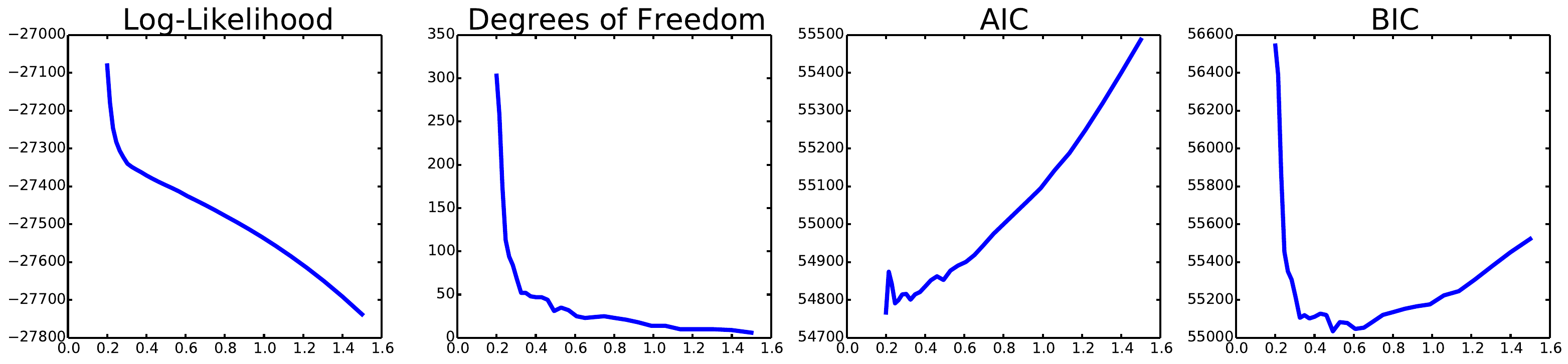}
\end{center}
\caption{\label{fig:solution_path} Left two panels: the log likelihood and degrees of freedom as the value of $\lambda$ changes from 1.5 to 0. Right two panels: corresponding AIC and BIC.  Empirically, AIC tends to lead to undersmoothing and worse FDR performance, but BIC finds a good balance point between fit and complexity.}
\end{figure}

Figure \ref{fig:solution_path} shows a typical solution path trace for the log likelihood, surrogate degrees of freedom, AIC, and BIC. In FDR smoothing problems, the number of plateaus is typically much smaller than the number of data points, and the penalty that AIC places on the degrees of freedom is dominated by the log likelihood. As a result, AIC is a disaster in practice, producing images that are far too grainy.  On the other hand, BIC achieves a much better balance across a range of problems, and we recommend it as a criterion for choosing $\lambda$.

An additional practical complication is that it is non-trivial to compute the number of plateaus efficiently for large-scale problems.  The na\"ive approach of counting the number of distinct values in $\hat{\boldsymbol\beta}$ can fail badly if the estimate has multiple spatially-separated plateaus with the same estimated prior probability (up to the precision of the ADMM convergence criterion). Pseudo-code for our plateau-counting method is provided in the appendix.

%% file: experiments.tex
\subsection{Setup and protocol}
\label{subsec:experiments:setup}
To demonstrate the effectiveness of FDR smoothing, we conducted simulation experiments across eight different scenarios defined by a full cross of three factors:
\begin{itemize}
\item two different configurations of the site-specific prior signal probability $c_i$ in a region of heightened signal probability. In one configuration, the true prior probability in this region is defined to be saturated at $c_i = 1$; that is, every site in the signal region is a signal. In the other configuration, the signal region is defined to be more mixed, with $c_i = 0.5$.
\item two different configurations of $c_i$ in the ``background'' region. We consider the cases where the background is purely from the null region, $c_i = 0$, and another where the background contains some signal, with $c_i = 0.05$. Figure \ref{fig:benchmark_example} (top left panel) shows an example of the site-specific prior distributions.
\item two different choices for $f_1(z)$, the distribution of test statistics under the alternative hypothesis. Each $f_1(z)$ is defined as the Gaussian convolution of some ``noiseless'' signal distribution $\pi(\theta)$. We consider a ``well-separated'' alternative, $\pi(\theta) = \frac{1}{2}\N(-2.5, 1) + \frac{1}{2}\N(2.5, 1)$, that has its modes far from the mode of the null distribution, and a ``poorly-separated'' alternative, $\pi(\theta) = \N(0, 3)$, that has the same mean as the null but with fatter tails.
\end{itemize}
In all eight scenarios, the spatial structure was a $128 \times 128$ two-dimensional grid graph, as in the fMRI example.  We simulated 30 data sets in each scenario and set the desired false discovery rate at $10\%$.

For each data set, we simulated $z$ scores as follows.  Let $\{c_i\}$ be the true image of prior probabilities, let $\pi(\theta)$ the true (noiseless) distribution of signals, and let $\delta_0$ be a Dirac measure at zero.  Then $z_i$ is drawn from the mixture model
\begin{eqnarray*}
\theta_i &\sim& c_i \pi(\theta) + (1-c_i) \delta_0 \\
z_i &\sim& \N(\theta_i, 1)  \, .
\end{eqnarray*}
The null hypothesis is that $\theta_i = 0$, in which case $z_i \sim f_0(z) =  \N(0,1)$.  The alternative hypothesis is that $\theta_i  \neq 0$, in which case $z_i$ is drawn from the Gaussian convolution $f_1(z) =  \int_\R \N(\theta_i,1) \ \pi(\theta_i) \ d \theta_i$.

\begin{figure}
\begin{center}
\includegraphics[width=0.45\textwidth]{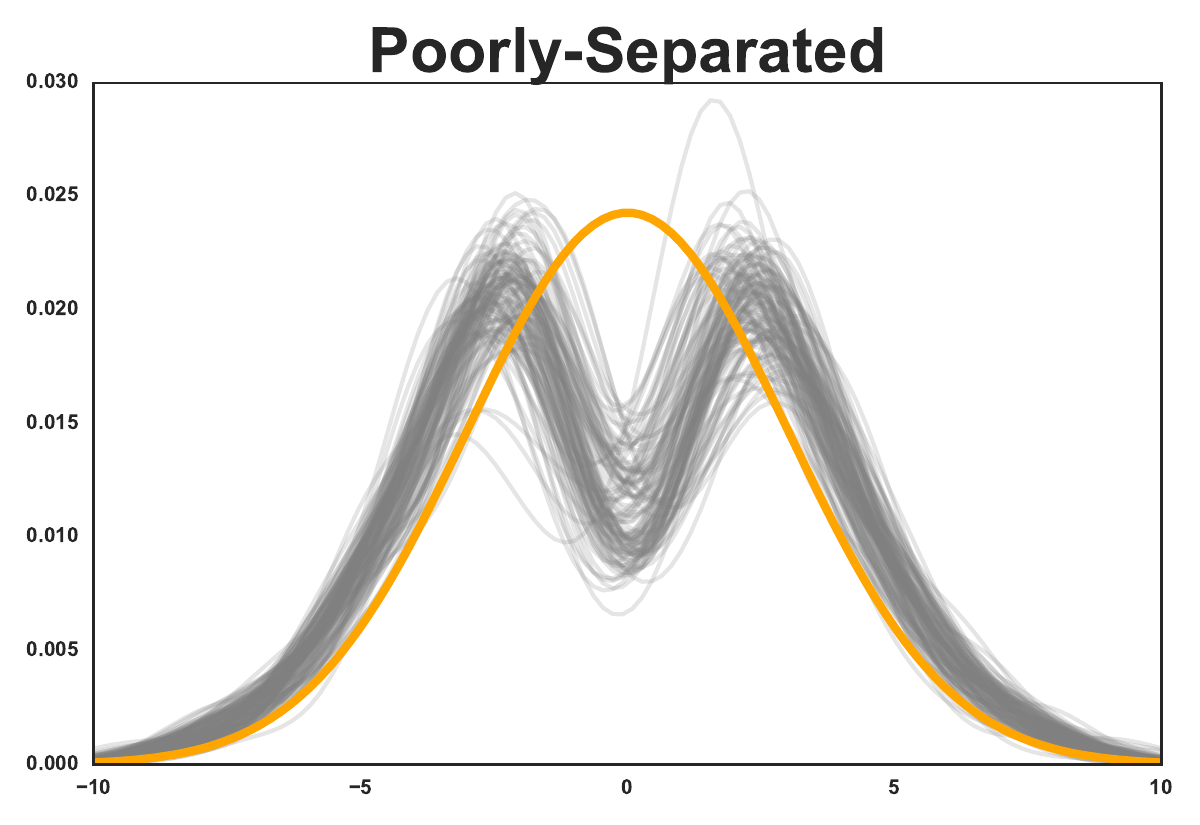}
\includegraphics[width=0.45\textwidth]{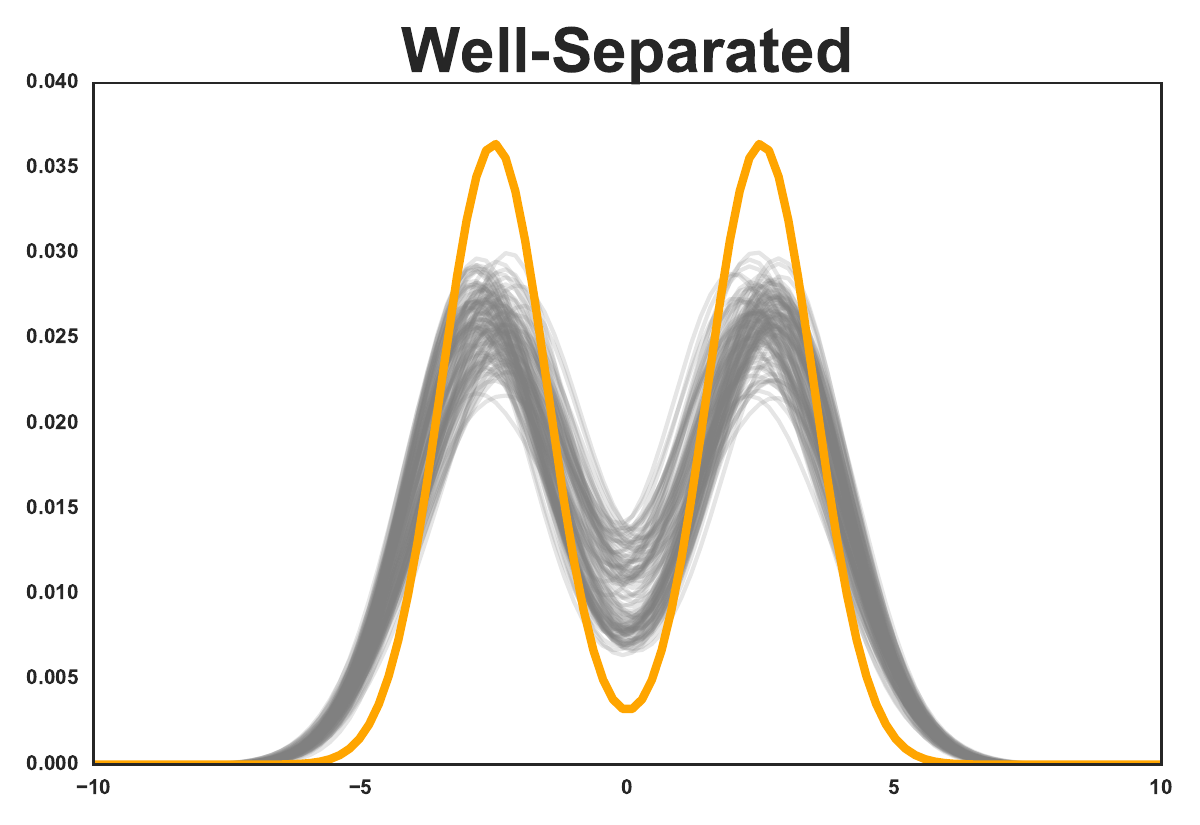}
\end{center}
\caption{\label{fig:estimated_signals} Visualization of the two $f_1(z)$ (alternative distribution) scenarios considered in the simulation experiments. The left panel shows the alternative distribution that is poorly separated from the standard-normal null distribution ($f_0(z)$), in the sense that they both have the same mode at 0.   The right shows the well-separated (bimodal) alternative distribution.  The thin gray lines in each panel show the resulting nonparametric estimates of $f_1(z)$ via predictive recursion for each of the 120 simulated data sets.}
\end{figure}

Figure \ref{fig:estimated_signals} shows the true true $f_1(z)$ (orange curve) in the two alternative hypothesis scenarios; the nonparametric estimates of $f_1(z)$ are shown in the thin gray lines. We show all 120 estimates from our benchmarks to convey a sense of the variance of the predictive recursion procedure, across a wide array of scenarios.


We compare our approach against three other techniques for multiple testing: 1) the Benjamini--Hochberg procedure \citep{benjamini1995}; 2) the $\text{FDR}_L$ procedure \citep{zhang:etal:2011}; and 3) the hidden Markov random field (HMRF) method of \citet{shu:etal:2015}. The first method is well known; we briefly explain the second and third. $\text{FDR}_L$ estimates the false discovery rate from locally smoothed $p$-values, obtained by taking the median of p-values in a local neighborhood on the graph. The HMRF method models the dependence of the site-specific priors using an Ising model.  The model is fit via an EM routine that relies on Gibbs sampling to compute the E step.  Finally, to establish a baseline we also present the results of an ``oracle'' two-groups model in which both the true $f_1(z)$ and the true underlying $\boldsymbol\beta$ vector are assumed known. The represents the theoretical limit of performance of the two-groups model. Details on how we ran $\text{FDR}_L$ and the HMRF model are provided in Appendix \ref{app:benchmark_setup}.

We also considered two spatial smoothing approaches, but ultimately did not include these are benchmarks. The first method, FDR-regression \citep{scott:kass:etal:2014}, requires introducing a large set of basis functions to model spatial dependence.  (It also requires that the underlying graph be embedded in a metric space.)  We did benchmark against FDR regression, but the resulting comparison added little to the overall presentation and insight: FDR-smoothing outperformed FDR regression soundly in all of our comparisons and we therefore chose not to include these results in the paper; these results are included in Appendix \ref{app:fdrregression} for completeness.  Another method \citep{sun:etal:2015} was determined to not be suitable for inclusion for several reasons: 1) it is focused on continuous spaces with a valid distance metric (our space has only an adjacency structure); 2) the reliance on testing an interval null of the form $H_0: |m(s)| < d$ for some sufficiently small $d$, which is not necessarily the same as testing a point null \citep{berger:delampady:1987}; and 3) the implementation of their method relies on using Gaussian processes.  In practice their code was not able to scale even to our synthetic dataset on a $128 \times 128$ grid, and certainly would be unable to scale to a full fMRI dataset of 750,000 voxels. We also note that although this benchmark example focuses on a single plateau of elevated prior probability, we conducted a suite of other experiments with varying plateau sizes, counts, and prior probabilities; we found that FDR smoothing performs similarly in all of these cases and thus chose to focus on the simplest experiment that still highlights the key differences among the available methods.


\subsection{Results and interpretation}
\label{subsec:experiments:results}
The benchmark results (Table \ref{tab:results}) reveal a failure of both \FDRL and HMRF: they are not robust to regions of interest which are not saturated (i.e.~100\% of the test statistics from $f_1$). Consequently, both methods substantially exceed the 10\% false discovery rate threshold (bottom sub-table; columns 3, 4, 7, and 8). FDR smoothing, in contrast, obeys the FDR threshold in all scenarios.

\begin{figure}
\begin{center}
\includegraphics[width=1.1\textwidth]{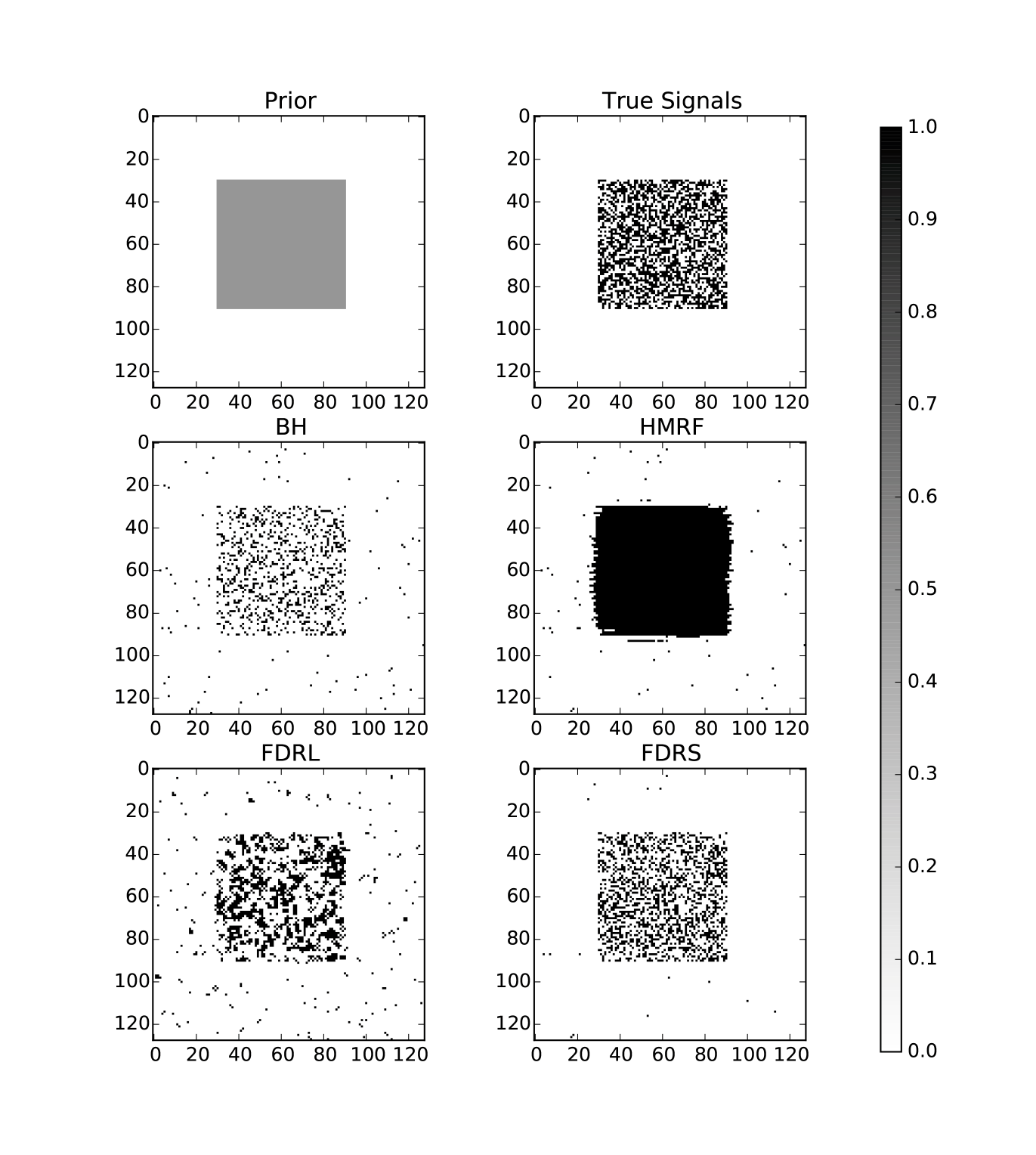}
\end{center}
\caption{\label{fig:benchmark_example} Results for a single trial from a scenario with a poorly-separated alternative, mixed signal region, and pure background region. Top left: the true priors ($\{c_i\}$) for all test sites. Top right: the test sites truly coming from the alternative distribution. Middle left: the signals detected by the Benjamini-Hochberg procedure. Middle right: the signals detected by the hidden Markov random field method. Bottom left: the signals detected by the $\text{FDR}_L$ procedure. Bottom right: the signals detected by our FDR smoothing method.}
\end{figure}

From a power perspective, FDR smoothing also provides clear advantages. In six of the eight examples, FDR smoothing has the highest true positive rate (TPR) among all methods which did not violate the $10\%$ FDR threshold. Furthermore, in each of these scenarios the FDR smoothing method was within 2-4\% of the oracle method, leaving little room for increased power. The only scenario in which FDR smoothing is slightly underpowered is when the signals are poorly separated and the signal regions are saturated. To better understand why HMRFs performed well in this scenario, we give a typical example of the results on a mixed signal region scenario in Figure \ref{fig:benchmark_example}. 

As shown in the upper right panel of Figure \ref{fig:benchmark_example}, the signals are clearly mixed within the signal region. Nevertheless, the HMRF model essentially estimates the entire signal region to be purely signal. This is a typical result for both mixed and saturated signal region scenarios. The difference is that under the saturated signal regime, estimating the entire signal region as 100\% signal happens to be the correct strategy. The implication here is that HMRFs are only preferable in the special case where the data contain plateaus of saturated signal with minimal noise in the other regions. However, in such cases, it seems likely that one could segment the data in a more effective manner by leveraging this prior knowledge. For additional details, see Appendix \ref{app:hmrf}.

\begin{table}
\begin{center}
\begin{tabular}{llllllllll}
\multicolumn{1}{l}{}  & \multicolumn{8}{c}{True positive rate (TPR)}                                                                    \\ 
\toprule
\multicolumn{1}{l}{$f_1$}  & \multicolumn{4}{c}{Well-Separated} && \multicolumn{4}{c}{Poorly-Separated}                         \\ 
\multicolumn{1}{l}{Signal Region}  & \multicolumn{2}{c}{Saturated} & \multicolumn{2}{c}{Mixed} && \multicolumn{2}{c}{Saturated} & \multicolumn{2}{c}{Mixed}                        \\ 
Background              & Pure & Noisy & Pure & Noisy && Pure & Noisy & Pure & Noisy\\
              \midrule
BH  &  0.500          & 0.516          & 0.416          & 0.451          && 0.415          & 0.428          & 0.370 & 0.388 \\
FDRL & 0.920          & 0.810          & 0.461          & 0.375          && 0.747          & 0.661          & 0.289 & 0.235 \\
HMRF & 0.990          & 0.924          & 0.996          & 0.795          && \textbf{0.989} & \textbf{0.910} & 0.926 & 0.553 \\
FDRS & \textbf{0.999} & \textbf{0.925} & \textbf{0.678} & \textbf{0.597} && 0.776          & 0.686          & \textbf{0.510} & \textbf{0.460} \\
\midrule
Oracle & 1.000 & 0.945 & 0.696 & 0.607 && 1.000 & 0.929 & 0.553 & 0.495\\
 
\\
\\
\multicolumn{1}{l}{}  & \multicolumn{8}{c}{False discovery rate (FDR)}                                                                    \\ 
\toprule
\multicolumn{1}{l}{$f_1$}  & \multicolumn{4}{c}{Well-Separated} && \multicolumn{4}{c}{Poorly-Separated}                         \\ 
\multicolumn{1}{l}{Signal Region}  & \multicolumn{2}{c}{Saturated} & \multicolumn{2}{c}{Mixed} && \multicolumn{2}{c}{Saturated} & \multicolumn{2}{c}{Mixed}                        \\ 
Background              & Pure & Noisy & Pure & Noisy && Pure & Noisy & Pure & Noisy\\
              \midrule
BH  &  0.079 & 0.073 & 0.089 & 0.085 && 0.078 & 0.074 & 0.087 & 0.085 \\
FDRL & 0.102 & 0.150 & 0.399 & 0.441 && 0.103 & 0.135 & 0.384 & 0.421 \\
HMRF & 0.097 & 0.054 & 0.550 & 0.452 && 0.102 & 0.047 & 0.481 & 0.254 \\
FDRS & 0.059 & 0.059 & 0.099 & 0.100 && 0.002 & 0.009 & 0.079 & 0.076 \\
\midrule
Oracle & 0.100 & 0.100 & 0.100 & 0.100 && 0.100 & 0.100 & 0.100 & 0.100\\
\end{tabular}
\caption{\label{tab:results} Results of the eight simulation studies. Each entry is an average error rate across 30 simulated data sets. The true-positive rates in bold are for the highest TPR model that does not violate the $10\%$ FDR threshold. FDR smoothing (FDRS) results in the highest admissible true-positive rate for all but two of the scenarios, consistently beating both the Benjamini--Hochberg procedure (BH) and FDRL. Crucially, FDR smoothing does not violate the FDR threshold for any of the experiments, whereas the two competing state-of-the-art methods substantially exceed the limit on all four mixed signal region simulations.}
\end{center}
\end{table}

The \FDRL procedure assumes that p-values are spatially dependent, with adjacent pixels having similarly-distributed p-values. Given this view of the problem, \FDRL aims to increase power by taking the median p-value of each test site and its neighbors.  It then corrects for multiplicity by appealing to facts about the order statistics of the uniform distribution. This modeling assumption is fundamentally different than in FDR smoothing, which assumes that the dependence lies in the prior weights of the two-groups model. As can be seen in the bottom left panel of Figure \ref{fig:benchmark_example}, the consequence of the \FDRL approach is that discoveries tend to clump together into small local clusters. This results in overestimating the number of discoveries in high-signal-density regions, as well as being sensitive to outlier neighborhoods.

Overall, FDR smoothing strikes an appealing balance of reliability, power, and speed, especially relative to the other available methods. Across a variety of experimental settings, FDR smoothing strictly adheres to the FDR threshold, providing important reassurances to the scientist.  Finally, all of our FDR smoothing scenarios could easily have been run on a laptop. Running the full solution path for an example like the one shown in Figure \ref{fig:benchmark_example} takes 14 minutes on a single node in our cluster environment, with 1/3 of that time spent performing 50 sweeps of predictive recursion; in practice, 5 sweeps is generally sufficient and would speed up FDR smoothing even more. For comparison, the same example takes over two days to run the HMRF model.


\subsection{Robustness to Misspecification}
\label{subsec:experiments:misspecification}
The two groups model we rely on makes several assumptions about the distribution of the observed z-scores. We investigate the robustness of FDR smoothing under the relaxation of two of these assumptions: a spatially-invariant alternative distribution and uncorrelated errors. At their core, these experiments are designed to see how well our algorithm performs when the z-scores are not conditionally independent as specified by the two groups model, but rather are related to their location and the values of their neighbors.

\begin{figure}
\hspace{-4mm}\makebox[\textwidth][c]{
\includegraphics[width=2.5in,height=2in]{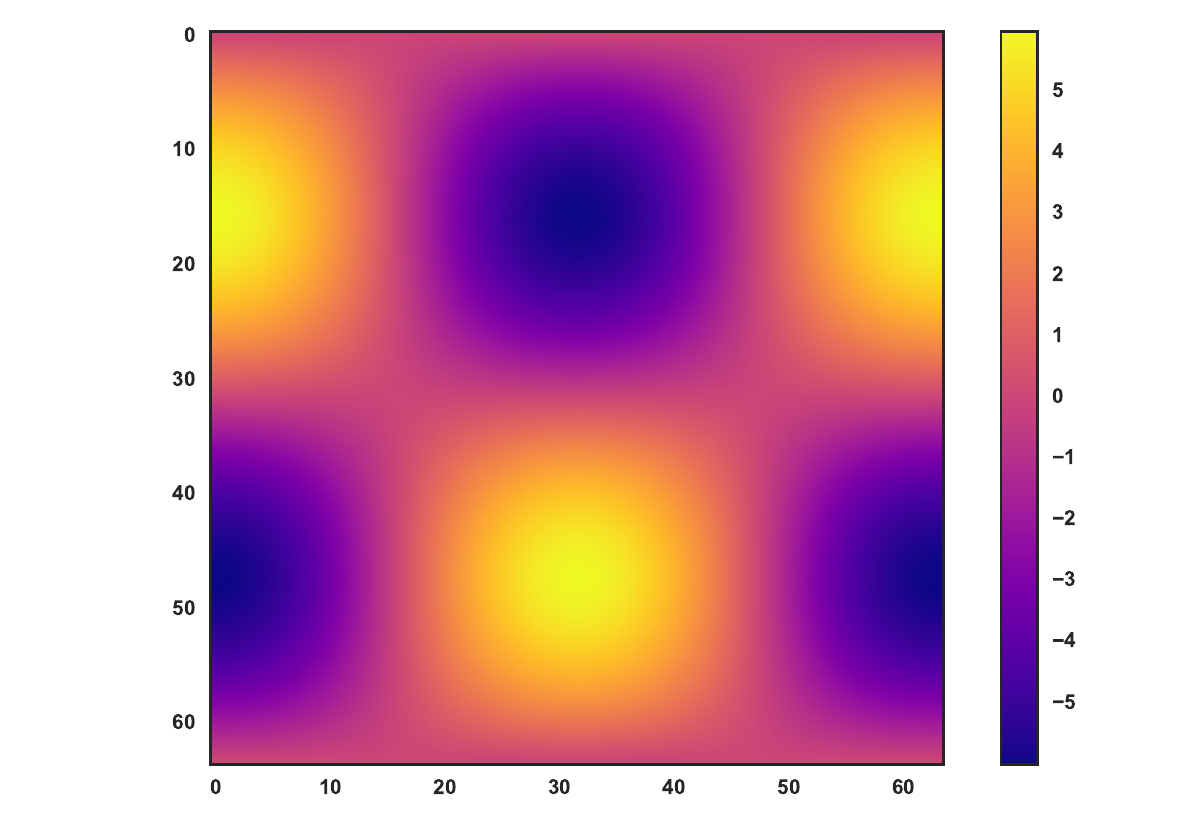}
\includegraphics[width=2in,height=2in]{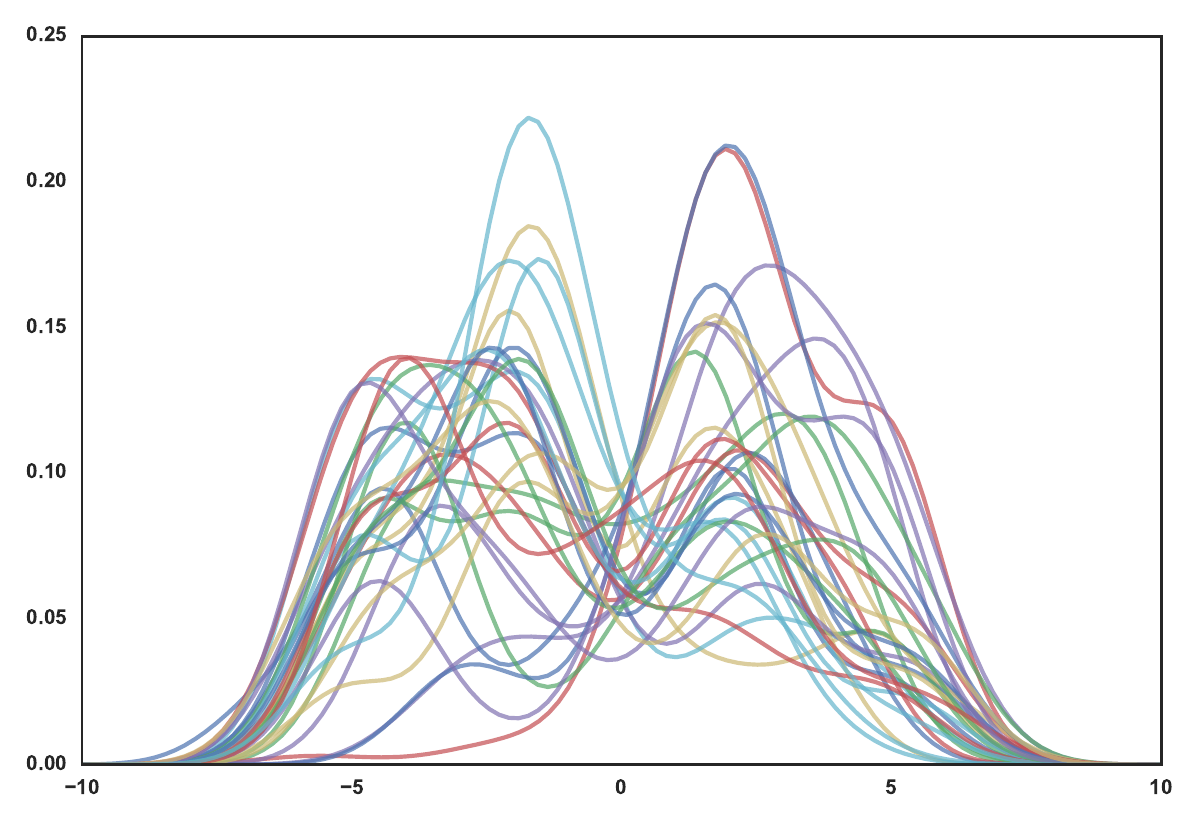}
\includegraphics[width=2in,height=2in,trim={0 1.5cm 0 1.5cm},clip]{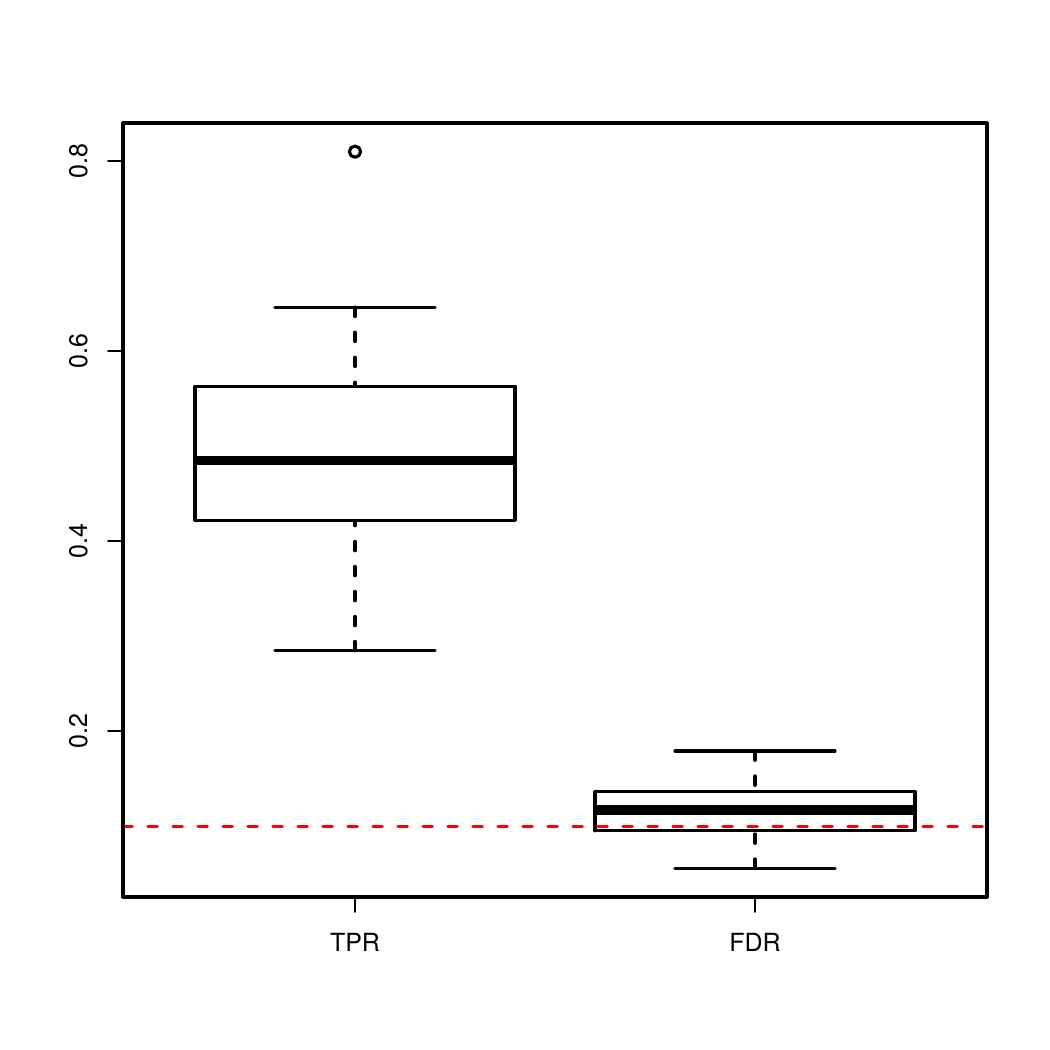}
}
\caption{\label{fig:spatial_z_performance} Left: The mean of the spatially-dependent alternative distribution in the first misspecification example. Middle: The global alternative distribution estimated in each trial by the predictive recursion procedure in FDR smoothing. Right: TPR and FDR for FDR smoothing in the spatially-dependent z-value scenario; FDR smoothing only slightly exceeds the 10\% FDR threshold (by approximately $1\%$ on average).}
\end{figure}

The alternative distribution estimated by our predictive recursion routine is assumed to be globally fixed and spatially invariant. However, it is reasonable to believe that some experiments may have a spatially-dependent alternative distribution. For instance, a significant difference in some region of the brain may be due to inhibition of neuronal activity while another region may experience excitation. To examine the performance of FDR smoothing under a spatially-dependent alternative distribution, we conducted 30 independent trials on a $64\times64$ grid where the alternative distribution was $N(\mu_{ij}, 1)$ with a spatially-dependent mean function: $\mu_{ij} = 6sin(\frac{2i\pi}{64})cos(\frac{2j\pi}{64})$ (Figure \ref{fig:spatial_z_performance}, left panel; here $i$ and $j$ index position on the grid.) In each trial, we randomly generated three plateaus by sampling contiguous 1000-point regions with replacement, and drawing the test statistics at each site from the corresponding $N(\mu_{ij}, 1)$ distribution. Figure \ref{fig:spatial_z_performance} (Middle) shows the result of each globally-estimated alternative distribution. In many trials, the model incorrectly estimates the alternative distribution to be a roughly-unimodal distribution centered near one of the two peak activation levels (-6 and 6) of the true distribution. Nonetheless, as Figure \ref{fig:spatial_z_performance} (Right) shows, the model maintains strong performance by achieving high power while only slightly exceeding the 10\% FDR threshold on average (mean FDR: 11.54\%).

Another fundamental assumption of our model is that each z-score is an independent draw from its corresponding mixture component.  We investigate our method's robustness to violations of this assumption, by drawing $z$-scores from a multivariate Gaussian process, $N(\boldsymbol\mu, \Sigma)$ defined as follows. For noise-only nodes, $\mu_i$ was zero, and for signal nodes we drew $\mu_i \sim \frac{1}{2}\delta_{-2} + \frac{1}{2}\delta_{2}$, where $\delta$ is a Dirac measure. The covariance matrix $\Sigma$ was that of a Gaussian process with a squared-exponential kernel $k(x_i, x_j) = \exp(\frac{1}{2} b^{-2} \vnorm{x_1 - x_2}^2 + \epsilon * \mathcal{I}[x_i = x_j])$, where $b$ is the bandwidth parameter, $\mathcal{I}$ is the indicator function, and $\epsilon$ is a small nugget term added for numerical stability. We generated signal regions as in the previous experiment (1000-point plateaus of connected vertices) and varied the strength of covariance between neighbors on the graph by increasing the bandwidth parameter. For each value of $b$, we simulated 10 independent data sets. As shown in Figure \ref{fig:correlated_errors_and_pathological} (left), the FDR smoothing model maintains the 10\% FDR threshold until the bandwidth exceeds $b\approx1$. We note that a bandwidth above 1 here produces visibly ``clumpy'' regions of correlated noise.  This would be easy to detect in real data, and is well beyond what one might expect in a low-to-moderate correlation setting. See Appendix \ref{app:correlated_noise} for an example visualization of the data at the first bandwidth above the $b=1$ threshold.


The FDR smoothing approach appears overall to be very robust to model misspecification.  In most situations, mis-specification leads to an increasingly conservative estimation of the discoveries, rather than by substantially exceeding the FDR threshold. It is only in the presence of noise with visible (and easily detected) spatial correlation that the model begins to violate the FDR threshold, which happens gradually as the spatial correlation increases. These results should give further confidence to the practitioner: using FDR smoothing may be useful even in scenarios where the assumed model does not completely fit the experimental design.

\subsection{Pathological Scenarios}
\label{subsec:experiments:pathological}
The one pathological example we have found where FDR smoothing fails catastrophically is when the number of test sites drawn from the alternative distribution is nearly as large, or larger, than the amount drawn from the null distribution. When this happens, the predictive recursion procedure may actually interpret the alternative distribution as the null and vice versa. The result is a complete inversion of the estimated prior regions and corresponding posteriors.

\begin{figure}
\vspace{-0.35in}
\centering
\begin{subfigure}{0.46\textwidth}
\includegraphics[width=\textwidth]{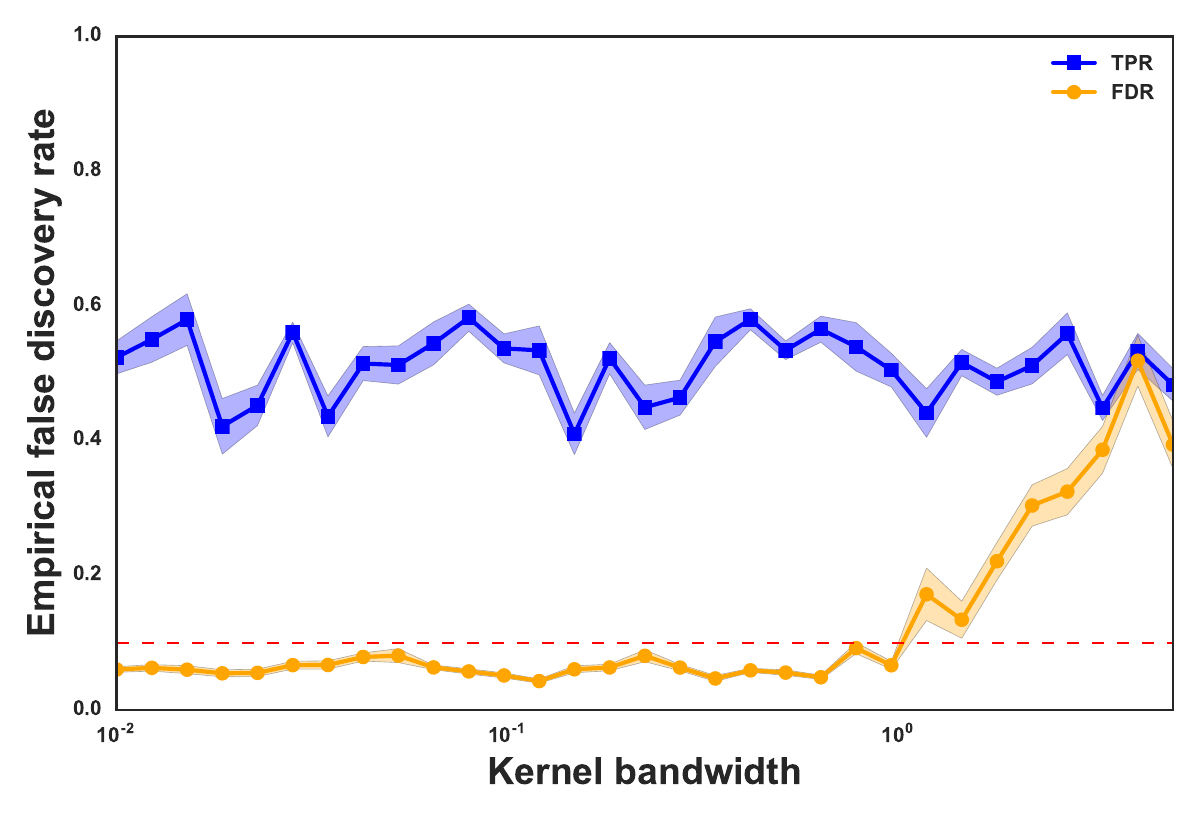}
\label{fig:correlated_errors_performance}
\end{subfigure}\qquad
\begin{subfigure}{0.46\textwidth}
\includegraphics[width=\textwidth]{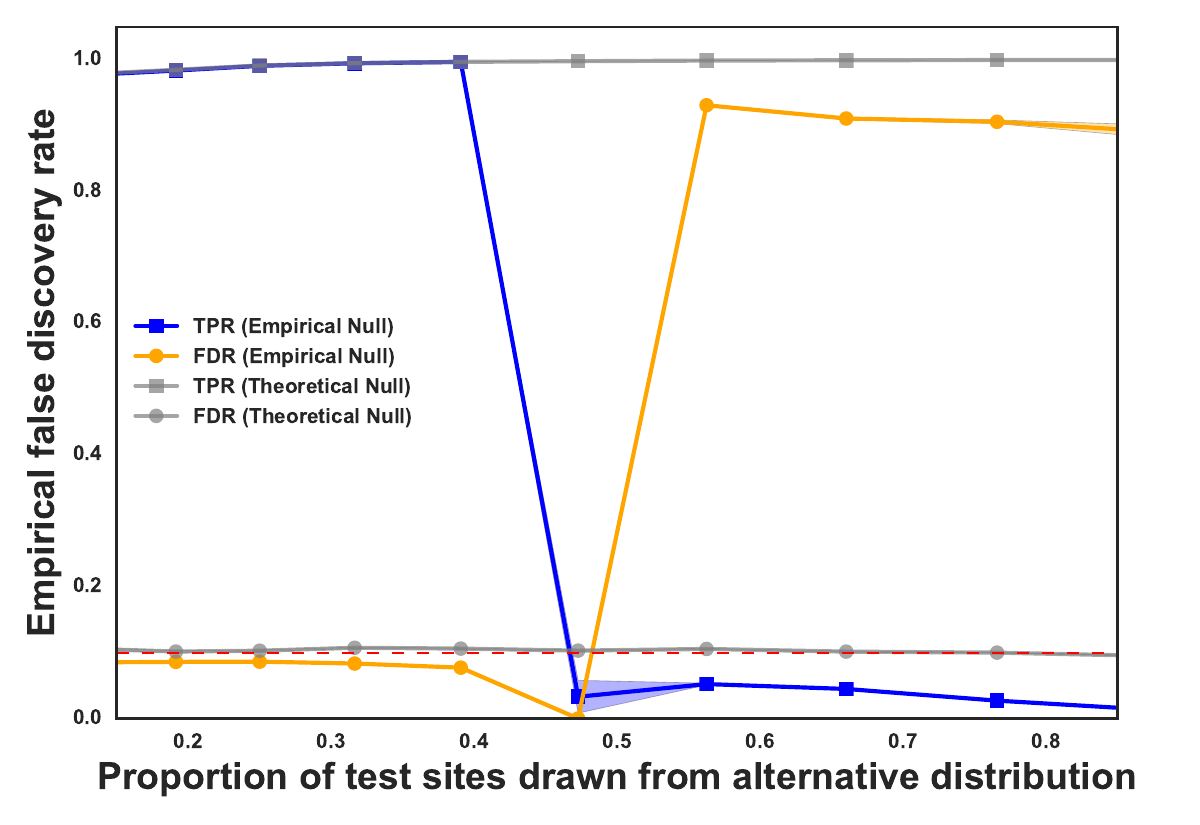}
\label{fig:pathological}
\end{subfigure}
\vspace{-0.3in}
\caption{\label{fig:correlated_errors_and_pathological} Left: The performance of the FDR smoothing algorithm when errors are correlated. Points correspond to means over 10 independent trials, with bands corresponding to standard error. The FDR threshold is conserved until the bandwidth reaches a pathologically high level. Right: The performance of the FDR smoothing algorithm as the number of test sites drawn from the alternative distribution increases. When more than half of the data is drawn from the alternative, the predictive recursion routine estimates the alternative distribution as the null distribution, and vice versa. The result is an inverted set of predictions, with mostly null points estimated to be discoveries.}
\end{figure}

To illustrate this point, we simulated along a continuum from entirely-null to entirely-alternative distribution samples. At each point on the grid, we ran 10 independent trials and measured TPR and FDR for the model. The null distribution was a standard normal and the alternative distribution was a normal distribution centered at 2. Figure \ref{fig:correlated_errors_and_pathological} (right) shows the performance of FDR smoothing as the proportion of test sites drawn from the alternative distribution varies. After the pathological threshold of \texttildelow50\% is crossed, we see that the TPR and FDR ``flip'', as the model now estimates all null sites to be discoveries.

Fortunately, this pathological scenario is easy to detect in practice. As a sanity check, one should simply examine the estimated null and alternative distributions. The purpose of the empirical null procedure is to correct for slight systematic bias in the experimental procedure that generated the data. One should therefore expect the resulting null estimation to be reasonably close to a standard normal. In the pathological case above, however, the empirical mean of the null distribution is very far from zero; similarly, the alternative distribution is almost precisely a standard normal. If this scenario arises, we advise the practitioner to simply use the theoretical null, as the benefit from the empirical null is much smaller than from the spatial smoothing procedure of FDR smoothing.

%% file: discussion.tex
Modern scientific analyses often involve large-scale multiple hypothesis testing. In many cases, such as fMRI experiments, these analyses exhibit spatial structure that is ignored by traditional methods for multiplicity adjustment.  As we have shown, exploiting this spatial structure via FDR smoothing offers a simple way to increase the overall power of an experiment while maintaining control of the false discovery rate.  Our method achieves this performance by automatically identifying spatial regions where the local fraction of signals is enriched versus the background.

While our results show strong statistical and computational performance, there are many areas in which our approach could be improved.  We call attention to a few of these areas and suggest them as subjects for future research.

First, the choice of a constant $\ell^1$ penalty on the first differences across edges in the graph leads to some slight overshrinkage in the estimated prior probabilities.  This is most evident in Figure \ref{fig:toy_example_1D}, where the estimated $c_i$ are shrunk back to the grand mean (or equivalently, toward the estimate of the ordinary two-groups model) versus the true $c_i$.  This reflects the well-known ``non-diminishing bias'' feature of the $\ell^1$ or lasso penalty, and is often mitigated in linear regression by using the adaptive lasso \citep{zou:2006} or a concave penalty \citep{mazumder:friedman:hastie:2009,polson:scott:2010b}.  Translating these ideas to the FDR smoothing problem presents a formidable algorithmic challenge and is an important area for future work.  Nevertheless, even with this noticeable overshrinkage, FDR smoothing achieves state-of-the-art performance in our synthetic experiments.  Moreover, it is possible that the overshrinkage is a feature rather than a deficit, in that it prevents the method from being too aggressive in hunting for very small regions of signals.


Second, our method for choosing $\lambda$, the regularization parameter, is effective but \textit{ad hoc}.  Our path-based approach would benefit from new theory on the degrees of freedom of the generalized lasso in mixture-model settings, or from an entirely different principle for choosing the tuning parameter in sparse estimation.

Third, we have presented FDR smoothing as a general method and provided examples that suggest its wide potential for application.  Perhaps the most obvious area in which it could be useful is in the analysis of fMRI data.  The literature on fMRI data analysis is large and mature, including the literature on multiplicity correction \citep[e.g.][]{hayasaka:nichols:2003,poldrack:2007a,nichols:2012}.  We have not attempted to benchmark FDR smoothing against some of these specialized methods, which exploit specific features of fMRI problems that may not hold more generally.  This comparison would be out of place in a paper intended for a general statistical audience, but we intend to pursue it in our future work.

Finally, both the lasso and the two-groups model sit (independently of one another) at the center of a large body of theoretical work.  We cannot hope to summarize this literature and merely refer to reader to \citet{bickel:ritov:tysbakov:2008} for the lasso and \citet{bogdan:ghosh:2008b} for the two-groups model.  Combining these two lines of work to produce a theoretical analysis of FDR smoothing represents a major research effort and is beyond the scope of this paper.  Nonetheless, given the strong empirical performance of the method, we are hopeful that such an analysis will someday bear fruit.

All code for FDR smoothing is publicly available in Python and R at \url{https://github.com/tansey/smoothfdr}.

%% file: fmri_data.tex
\section{Details of fMRI data set}
\label{app:fmri_details}

The fMRI data set analyzed in Section 3 was acquired and processed as follows.  A spatial working memory localizer \citep{fedo:dunc:kanw:2013} was performed by a single subject.  On each trial, a 4x2 spatial grid is presented, and locations in that grid are presented sequentially (1000 ms per location), followed by a forced-choice probe between two grids, one of which contained all of the locations presented in the preceding series.  In the easy condition, one location is presented on each presentation, whereas in the hard condition two locations are presented on each presentation.  Twelve 32-second experimental blocks were interspersed with 4 16-second fixation blocks (acquisition time = 7:28).  The contrast presented in Figure 1 compares the hard versus easy conditions.

fMRI acquisition was peformed using a multi-band EPI (MBEPI) sequence \citep{moel:yaco:olma:2010} (TR=1.16 ms, TE = 30 ms, flip angle = 63 degrees, voxel size = 2.4 mm X 2.4 mm X 2 mm, distance factor=20\%, 64 slices, oriented 30 degrees back from AC/PC, 96x96 matrix, 230 mm FOV, MB factor=4, 10:00 scan length).   fMRI data were preprocessed according to a pipeline developed at Washington University, St. Louis \citep{powe:mitr:laum:2014}, including realignment for motion correction, distortion correction using a field map, and registration to a 3-mm isotropic atlas space.  Preprocessed task fMRI data were analyzed at the first level using the FSL Expert Analysis Tool (FEAT, version 5.0.6), using prewhitening and high-pass temporal filtering (100 second cutoff). 

%% file: plateaus.tex
\section{Finding plateaus in 2D images}
\label{app:plateaus}

Algorithm \ref{alg:plateaus} outlines our approach to finding plateaus, which is needed in our path-based algorithm for choosing $\lambda$. Note that each point in the grid is touched at most $k$ times, where $k$ is the number of neighbors of that point. Thus the algorithm runs in $\mathcal{O}(kn)$, which is effectively linear time since $k \ll n$.  The algorithm is mildly sensitive to underlying numerical inaccuracies in the ADMM solution for $\beta$.  It is well known that finite-precision ADMM solutions tend to slightly ``round off'' sharp edges in the underlying image.  This produces some slight numerical noise in the degrees of freedom estimate.  In our experience, this is rarely a practical concern, and can always be corrected by tightening the convergence criterion for ADMM below the plateau tolerance in Algorithm \ref{alg:plateaus}.

\begin{algorithm}[t]
\caption{Our plateau-finding algorithm.}
\label{alg:plateaus}
\begin{algorithmic}[1]
    \Require grid of values $\beta$, plateau tolerance $\epsilon$
    \Ensure list of plateaus and their values $\phi$
    \State $tocheck \gets coordinates(\beta)$
    \State $checked \gets \{\emptyset\}$
    \State $\phi \gets \{\emptyset\}$
    \While{$tocheck$ not empty}
        \State $(x_0,y_0)\gets$  pop $tocheck$ until $(x_0, y_0) \not\in checked$
        \State $points\gets \{(x_0,y_0)\}$
        \State $\beta_{min}, \beta_{max} \gets \beta_{x_0,y_0} - \epsilon, \beta_{x_0,y_0} + \epsilon$
        \State $unchecked\gets \{(x_0, y_0)\}$
        \While{$unchecked$ not empty}
            \State $(x,y) \gets $ pop $unchecked$
            \ForAll{neighbor $(v,w)$ of $(x,y)$}
                \If $(v,w) \not\in checked$ and $\beta_{min} \leq \beta_{v,w} \leq \beta_{max}$
                    \State Add $(v,w)$ to $points$, $unchecked$, and $checked$
                \EndIf
            \EndFor
        \EndWhile
        \State Add $points$ to $\phi$
    \EndWhile
    \State
    \Return $\phi$
\end{algorithmic}
\end{algorithm}

%% file: benchmark_setup.tex
\section{Benchmark setup}
\label{app:benchmark_setup}

As described in Section \ref{sec:experiments}, all methods were run across a suite of scenarios, with 30 independent trials per scenario and a 10\% FDR threshold. This appendix describes the method-specific settings for the two main competing methods: \FDRL and the HMRF model.

For \FDRL, we used ``Method 1'' from \citep{zhang:etal:2011} as this was suggested for fMRI-type data. We set the null-cutoff $\lambda = 0.2$. This is higher than used in \citep{zhang:etal:2011}, which used $\lambda = 0.1$; however, they also used a 1\% FDR threshold. Since $\lambda$ controls the proportion above which we expect almost all p-values to be true nulls, using a $\lambda$ of $0.2$ is more reasonable with an FDR of $0.1$. Preliminary experiments confirmed the \FDRL authors' claim that \FDRL is not very sensitive to the setting of $\lambda$.

The HMRF model has several tunable parameters and required tweaks to run the code provided in the supplementary materials of \citep{shu:etal:2015}:

\begin{itemize}
\item In order to compile the C++ code, we needed to change all calls to \texttt{floor(x)} with \texttt{(double((int)x))}.
\item 2d grids and edge points are not supported in their implementation. To process the entire $128\times128$ grid, we had to embed it within the center of a $3\times130\times130$ array. This should have no effect on the result, as we specified the original lattice as the region of interest.
\item The alternative density estimation procedure is parametric (as opposed to our nonparametric approach) and requires specifying the number of components in a Gaussian mixture model. We specify the correct number of components in each case, to give their model the best possible estimation (i.e. 2 for the well-separated scenarios and 1 for the poorly-separated scenarios).
\item We ran with the default parameters of $sweep_b = 1000$, $sweep_r = 5000$, $sweep_lis = 1e6$, $iter_max = 5000$. These correspond to 5000 iterations of the main Gibbs sampler with a 1000-iteration burn-in. These settings are identical to those used in the HMRF paper.
\end{itemize}

We made every effort possible to be as generous as possible to both methods. This is the main reason for choosing to include the ``saturated'' signal regions, as these cases highlight the areas where \FDRL and HMRFs perform well, even though we expect them to be rare in practice, as evidenced by the various prior plateaus discovered by FDR smoothing in Figure \ref{fig:russ_example_slice}.

%% file: fdrregression.tex
\section{Comparisons with FDR Regression}
\label{app:fdrregression}

Benchmark performance results against FDR regression (FDR-R) are presented in Table \ref{tab:fdrregression_results}. We performed 100 independent trials for each of eight different scenarios, corresponding to two different plateau setups (large regions vs. small regions) and the following four different alternative distributions:

\begin{enumerate}
\item $0.48 N(-2, 1) + 0.04 N(0, 16) + 0.48 N(2, 1)$
\item $0.4 N(-1.25, 2) + 0.2 N(0, 4) + 0.4 N(1.25, 2)$
\item $0.3 N(0, 0.1) + 0.4 N(0, 1) + 0.3 N(0, 9)$
\item $0.2 N(-3, 0.01) + 0.3 N(-1.5, 0.01) + 0.3 N(1.5, 0.01) + 0.2 N(3, 0.01)$
\end{enumerate}

FDR regression using a 100-dimensional $b$-spline basis comes close to the performance of FDR smoothing, but also has many conceptual and computational disadvantages. These are essentially the same disadvantages that one would face in treating \textit{any} spatial smoothing problem in a regression framework. For example, to handle a smoothing problem using FDR regression, one must choose the basis set and the number of basis elements. This is implicitly a choice about the smoothness of the underlying prior image, and is not straightforward in large problems or problems over an arbitrary graph structure. FDR smoothing, on the other hand, has no tunable parameters once our path-based method for choosing $\lambda$ is used.   Moreover, FDR regression cannot localize sharp edges in the underlying image of prior probabilities, unless these edges happen to coincide with any edges present in the basis set.  FDR smoothing finds these edges automatically without requiring a clever choice of basis, and without having to tolerate undersmoothing in other parts of the image. Finally, at an algorithmic level, the important matrix operations in FDR smoothing involve very sparse matrices and benefit enormously from pre-caching. This is not true in FDR regression, which involves dense matrices and linear systems that change at every iteration.

As the table shows, FDR regression with basis functions does provide sensible answers and good FDR performance.  However, the FDR smoothing approach benefits greatly by exploiting the spatial structure of the problem, resulting in better power and more interpretable summaries at lower computational cost.

\begin{table}
\begin{center}
\begin{tabular}{llllllllll}
\multicolumn{1}{l}{}  & \multicolumn{9}{c}{True positive rate (TPR)}                                                                    \\ 
\toprule
\multicolumn{1}{l}{}  & \multicolumn{4}{c}{Large Regions} && \multicolumn{4}{c}{Small Regions}                         \\ 
              & Alt 1 & Alt 2 & Alt 3 & Alt 4 && Alt 1 & Alt 2 & Alt 3 & Alt 4\\
              \midrule
BH  & 0.364   & 0.215   & 0.128   & 0.366   && 0.212   & 0.123   & 0.090   & 0.194 \\
2G  & 0.394   & 0.229   & 0.134   & 0.403   && 0.211   & 0.123   & 0.091   & 0.196 \\
FDR-R & 0.559   & 0.334   & 0.167   & 0.610   && 0.242   & 0.141   & \textbf{0.097}   & 0.232 \\
FDRS   & \textbf{0.592}   & \textbf{0.352}   & \textbf{0.168}   & \textbf{0.645}   && \textbf{0.264}   & \textbf{0.144}   & 0.093   & \textbf{0.257} \\
\midrule
Oracle & 0.688 & 0.524 & 0.332 & 0.718 && 0.298 & 0.193 & 0.139 & 0.292\\
 
\\
\\
\multicolumn{1}{l}{}  & \multicolumn{9}{c}{False discovery rate (FDR)}                                                                    \\ 
\toprule
\multicolumn{1}{l}{}  & \multicolumn{4}{c}{Large Regions} && \multicolumn{4}{c}{Small Regions}                         \\ 
              & Alt 1 & Alt 2 & Alt 3 & Alt 4 && Alt 1 & Alt 2 & Alt 3 & Alt 4                                        \\
              \midrule
BH  & 0.072   & 0.070   & 0.073   & 0.070   && 0.090   & 0.093   & 0.093   & 0.092 \\
2G  & 0.089   & 0.083   & 0.083   & 0.089   && 0.092   & 0.096   & 0.098   & 0.096 \\
FDR-R & 0.075   & 0.058   & 0.050   & 0.086   && 0.102   & 0.106   & 0.109   & 0.105 \\
FDRS   & 0.072   & 0.057   & 0.054   & 0.079   && 0.092   & 0.095   & 0.098   & 0.096 \\
\midrule
Oracle & 0.101 & 0.100 & 0.100 & 0.101 && 0.097 & 0.101 & 0.101 & 0.098 \\
\end{tabular}
\caption{\label{tab:fdrregression_results} Results of the eight simulation studies. Each entry is an average error rate across 100 simulated data sets.  FDR smoothing (FDRS) results in the highest true-positive rate for all but one of the scenarios, consistently beating both the Benjamini--Hochberg procedure (BH) and the two-groups model (2G).  FDR regression (FDR-R) comes close, but slightly overshoots the desired FDR limit of 10\% in the small-signal examples.    \citep{scott:kass:etal:2014} also report this behavior.  In contrast, FDR smoothing remains (on average) under the nominal FDR across all experiments.}
\end{center}
\end{table}

%% file: hmrf.tex
\section{HMRF details and improvements}
\label{app:hmrf}

\begin{figure}
\begin{center}
\includegraphics[trim={2cm 0 0 0},clip]{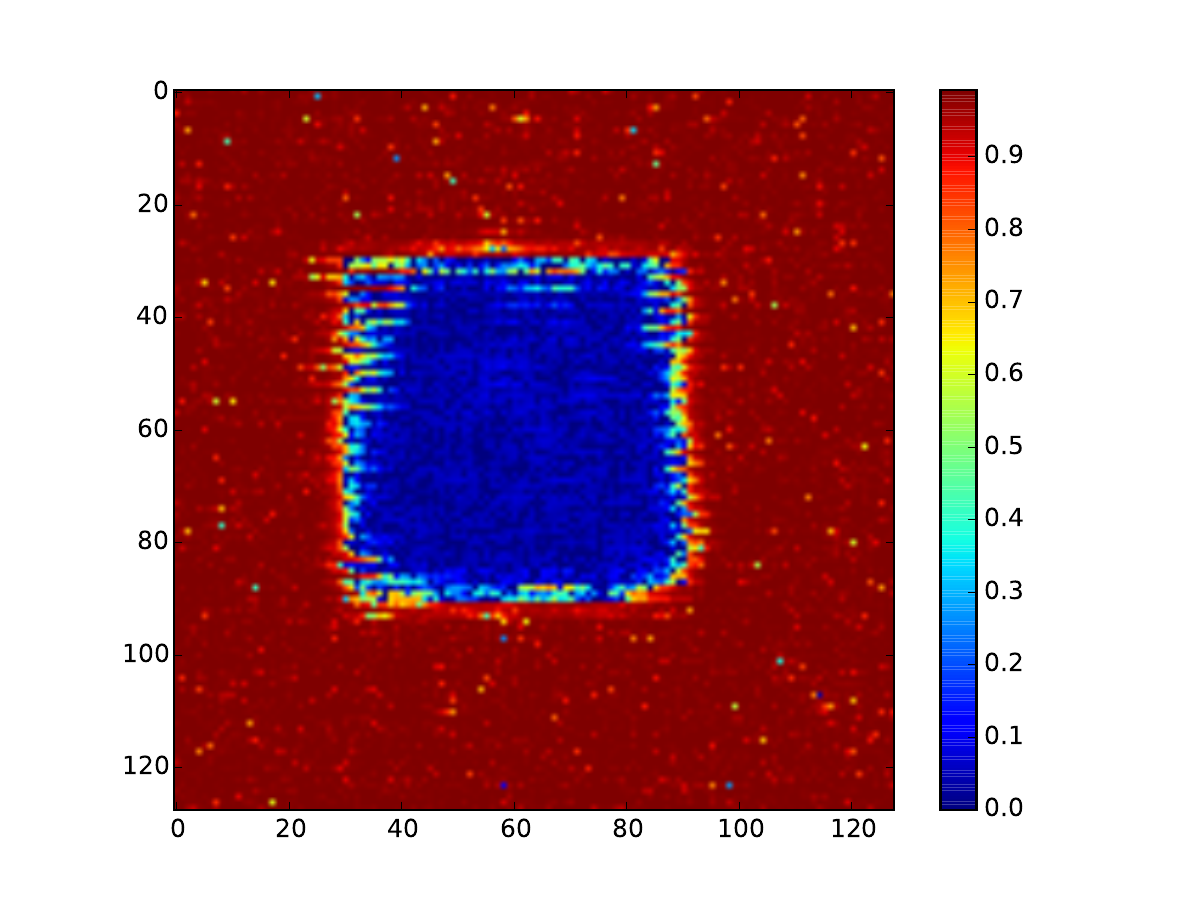}
\end{center}
\caption{\label{fig:hmrf_lis} The inferred ``local index of significance'' statistics inferred by the HMRF model on the example in Figure \ref{fig:benchmark_example}. The Ising model assumption, combined with the difficult-to-fit distribution it induces, results in a model that overestimates the strength of the signal region.}
\end{figure}

The HMRF model, while following the prior-dependence philosophy of FDR smoothing, makes a different distributional assumption on the dependence by placing an Ising model on the priors. This has two important side effects. First, the model is not necessarily going to discover constant regions of prior probability. This is clear when looking at the ``local index of significance'' (LIS) statistics produced by the HMRF, shown in Figure \ref{fig:hmrf_lis}. While the LIS space is substantially smoothed, it is not constant across different plateaus like in FDR smoothing. The other core issue with the HMRF model is that its substantial complexity results in a very difficult model to fit. The implementation provided by the authors performs an EM algorithm with Gibbs sampling and required more than three days to run the examples with the suggested number of iterations, compared to minutes with FDR smoothing on the same examples and the same compute cluster. More to the point, the final fit shows clear bias to local optima that over-estimate the strength of the signal region. The result is a model which performs well only when the regions are clearly segmented and the signal region is saturated, and which otherwise fails to adhere to the specified FDR threshold. See Appendix \ref{app:hmrf} for more details on the HMRF model, its configuration, and suggestions from the HRMF authors on ways to improve the runtime and fit of the model; we did not incorporate these suggestions in our benchmarks as they were either purely computational speedups or were intuitive suggestions that would require developing entirely new methods.

In an effort to provide fair evaluation, we contacted the first and second authors of the HMRF paper \citep{shu:etal:2015}. They provided several suggestions for improving the speed of the algorithm and its performance. The following speedup suggestions were offered:

\begin{itemize}
\item Reduce the number of burn-in iterations.
\item Monitor the stability of the parameter estimations in order to stop earlier than the maximum number of iterations.
\item Stop the backtracking line search at a fixed number of steps in the Newton's method step.
\item Use an updated pseudo-random number generator as the code relies on an outdated generator which may be slower than the most up-to-date version.
\end{itemize}

Note that all of the above suggestions would reduce the running time of the algorithm, but would not likely result in an improved fit or better performance on the benchmarks. The main performance improvement suggested was to preprocess the z-scores so as to detect the different regions first, then run separate HMRFs on each region. One way to do this would be to run FDR smoothing, then treat each plateau as a different region and fit an HMRF to them. It is unclear whether this approach would truly address the underlying issues we observed in the benchmarks. Thus, while this is an interesting idea and may be effective, it would constitute an entirely new method and therefore we leave it to future work.

%% file: correlated_noise.tex
\section{Correlated noise example with large bandwidth}
\label{app:correlated_noise}

\begin{figure}
\centering
\includegraphics[height=0.8\textheight]{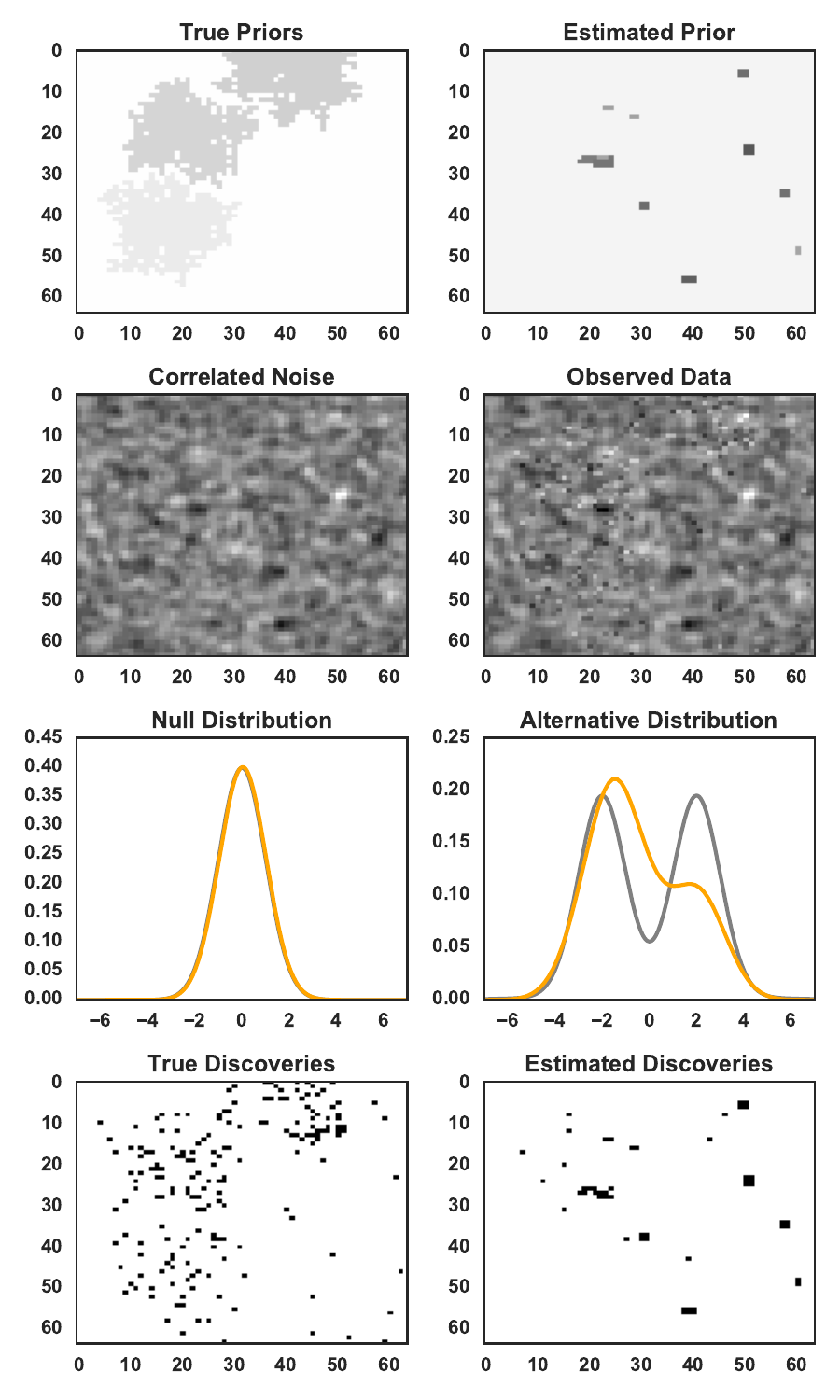}
\caption{\label{fig:correlated_noise_example} An example of a dataset generated with a bandwidth just greater than 1. The left figure in the second row shows the highly-correlated noise added to the model. The corresponding right figure shows the resulting data that the model is given, with clear examples of phantom plateaus.}
\end{figure}

Figure \ref{fig:correlated_noise_example} shows an example of a dataset from the experiment in Section \ref{subsec:experiments:misspecification}. The highly correlated noise creates clear regions of false positives that are difficult to distinguish from the true positive regions.

